\documentclass[aps,superscriptaddress,noshowpacs,preprintnumbers]{revtex4}

\usepackage{amsfonts}
\usepackage{varioref,exscale,latexsym,amsmath,amssymb,wasysym,soul}
\usepackage{graphicx,xcolor}
\usepackage{bm}
\usepackage{slashed}
\usepackage[colorlinks=true, pdfstartview=FitV, linkcolor=blue, citecolor=blue, urlcolor=blue]{hyperref}

\newcommand{\mockbar}{\mockbarsaturn}
\renewcommand{\mockbar}{{\mathord{\mathchoice
    {\mkern-1mu\text{\it\saturn}\mkern1.5mu}
    {\mkern-1mu\text{\it\saturn}\mkern1.5mu}
    {\mkern-.67mu\text{\scriptsize\it\saturn}\mkern1mu}
    {\mkern-.33mu\text{\tiny\it\saturn}\mkern.67mu}}}}

\newcommand{\be}{\begin{equation}}
\newcommand{\ee}{\end{equation}}
\newcommand{\beq}{\begin{eqnarray}}
\newcommand{\eeq}{\end{eqnarray}}
\newcommand{\bea}{\begin{eqnarray}}
\newcommand{\eea}{\end{eqnarray}}
\newcommand{\beqn}{\begin{eqnarray}}
\newcommand{\eeqn}{\end{eqnarray}}
\newcommand{\rd}{\mathrm{d}}
\newcommand{\rD}{\mathrm{D}}

\newcommand{\pa}{\partial}

\def\pa{\partial}

\def\pa{\partial}

\def\cO{\mathcal{O}}

\let\sss=\scriptscriptstyle
\parskip=\smallskipamount

\def\define{\mathrel{\overset{\sss\rm def}=}}

\begin{document}

\title{On the Emergent ``Quantum'' Theory in Complex Adaptive Systems}

\author{Tristan Hubsch}
\affiliation{Department of Physics and Astronomy, Howard University, Washington, D.C., 20059, U.S.A.}
\author{Djordje Minic}
\affiliation{Department of Physics, Virginia Tech, Blacksburg, VA 24061, U.S.A.}
\author{Konstantin Nikolic}
\affiliation{School of Computing and Engineering, University of West London, London W5 5RF, United Kingdom}
\author{Sinisa Pajevic}
\affiliation{Section on Critical Brain Dynamics, National Institute of Mental Health, NIH, Bethesda, MD 20892, U.S.A.}

\date{\today}

\begin{abstract}
We explore the concept of emergent quantum-like theory in complex adaptive systems, 
and examine in particular the concrete example of such an emergent (or ``mock'') quantum theory in the Lotka-Volterra system.
In general, we investigate the possibility of implementing the mathematical formalism of quantum mechanics on classical systems, and what would be the conditions for using such an approach. We start from a standard description of a classical system via Hamilton-Jacobi (HJ) equation and reduce it to an effective Schrodinger-type equation, with a (mock) Planck constant $\mockbar$, which is system-dependent. The condition for this is that the so-called quantum potential $V_Q$, which is state-dependent, is cancelled out by some additional term in the HJ equation. We consider this additional term to provide for the coupling of the classical system under consideration to the ‘environment’. We assume that a classical system could cancel out the $V_Q$ term (at least approximately) by fine tuning to the environment. This might provide a mechanism for establishing a stable, stationary states in (complex) adaptive systems, such as biological systems. 
In particular, we present a general argument as to why the non-equilibrium dynamics of a classical system could lead to a mock quantum description that ensures
stability compatible with adaptability.
In this context we emphasize the state dependent nature of the mock quantum dynamics and we also introduce the new concept of
the mock quantum, state dependent, statistical field theory.
We also discuss some universal features of the quantum-to-classical as well as the mock-quantum-to-classical transition
found in the turbulent phase of the hydrodynamic formulation of our proposal.
In this way we re-frame the concept of decoherence into the concept of ‘quantum turbulence’, i.e. that the transition between quantum and classical could be defined in analogy to the transition from laminar to turbulent flow in hydrodynamics.

\end{abstract}

\maketitle

\section{Introduction} 
\label{s:1}
The idea that new physics \cite{es} such as quantum theory has a fundamental role to play in living systems is a fascinating one, and has recently
captured the imagination of many physicists working in the new field of ``quantum biology'';
for a review and also for a critical view of this endeavor consult \cite{quantumbiology}, for example. 
But we note here that this is typically related to already established quantum phenomena --- just applied to biological systems. However, recently there has been a growing interest in using the equations of quantum mechanics for the explanation of experimental observations in nominally classical systems such as those found in neuroscience, psychology, economics etc --- sometimes the observed data do not conform with the laws of classical probability, but rather with quantum-like statistics \cite{plotnitsky,khrenikov}. This is the case when the underlying system is not necessarily quantum in its nature, but its behavior could be interpreted or predicted in the probabilistic sense using the mathematical formulation borrowed from quantum mechanics.
In this paper we extend our previous work \cite{Minic:2014zsa} on the emergent (``mock
,'' or analog) quantum theory in complex adaptive systems as a new viewpoint on 
the question whether or not new physics, such as quantum-like dynamics, could play an important role in the domain of living systems.

The concept of emergence is well known in condensed matter physics \cite{wen}. 
In this paper we revisit the concept of an emergent ``quantum'' framework (see also \cite{tony, droplet, fluidpilot, droplet2, Bush-Oza, eqbook}) which fits in the general notion of emergence \cite{anderson}.  Based on the unique properties of quantum systems \cite{qcqi}, such as enhanced stability and computational ability, we argue for an evolutionary advantage of the emergent ``quantum'' behavior in certain regimes, characterized by an emergent, non-universal and system-dependent, or mock, Planck constant, $\mockbar$. In particular, such non-classical stability and adaptability,
associated with this emergent quantum behavior, should be advantageous for complex adaptive systems. 

Herein, we adopt the notion of a ``complex adaptive system'' as a system composed of many components which typically affect one another via a dynamic network of interactions, and in such a way that the behavior of the whole system cannot be generically predicted from the behavior of its components. (This inherent irreducibility is a hallmark of the {\em emergent\/} behavior that we shall identify as ``mock-quantum.'') A complex system is adaptive in that the individual and collective behavior of the entire system either mutates or self-organizes corresponding to the interaction of the system and its environment, being able to adapt to the changing environment and increase the ability of the whole system to survive~\cite{jhholland}.

The general relevance of quantum-like behavior in biology was conjectured a long time ago \cite{bohr}, and the relation between non-linear dissipative systems and emergent quantum theory 
was explored more recently in \cite{thooft}.
Whereas quantum phenomena have been advocated to be important in the context of living systems \cite{quantumbiology, davies}, the usual concerns regarding decoherence and the difficulty of scaling-up of quantum phenomena in macroscopic systems are still the major conceptual obstacles to such discussions \cite{davies}.
This provides one of our key reasons to explore emergent (mock) quantum theory in this paper. Our main motivation is the ongoing experimental and theoretical work
as well as the ongoing critical discussion on
the emergent quantum-like physics found in the context of fluid dynamics \cite{droplet, fluidpilot, Bush-Oza, droplet_critique, droplet_critique2}. In this paper we provide a general 
discussion of mock quantum theory, but also examine it using a concrete example of 
the application of mock quantum theory to the Lotka-Volterra system.
In this context we emphasize the state dependence of the mock-quantum dynamics and we also introduce the new concept of
a mock-quantum, state-dependent, statistical field theory.
We also discuss some universal ``turbulent'' aspects of the quantum-to-classical and mock-quantum-to-classical transitions.

The paper is organized as follows:
In Section~\ref{s:2} we review the construction of Ref.~\cite{Minic:2014zsa}, and the de-Broglie-Bohm formulation and quantum potential of mock quantum theory in Section~\ref{s:3}.
In Section~\ref{s:4} we present the general argument regarding the question of ``why would the non-equilibrium dynamics of a classical system lead
to a mock quantum description''?
Section~\ref{s:5} summarizes our main example, the Lotka-Volterra system, and Section~\ref{s:6} examines the state-dependent dynamics of the mock quantum Lotka-Volterra system for two species.
Section~\ref{s:7} is devoted to the issue of stochasticity in our mock quantum dynamics via a mock-quantum statistical field theory.
In Section~\ref{s:8} we examine the robustness of the central aspect of our construction, the mock-quantum potential, in the context of maximal variety in complex systems. Section~\ref{s:9} is devoted to a hydrodynamic formulation of mock quantum theory and some universal ``turbulent'' features of the quantum-to-classical and mock-quantum-to-classical transitions. Finally, we present our conclusions in Section~\ref{s:10}. 
We also include Appendix~\ref{a:Hydro} dscribing the ``walking droplet'' system that exhibits quantum-like behavior, and two appendices, \ref{a:B} and~\ref{a:C}, collect some standard representations of quantum, and thus mock quantum theory.

\section{Emergent Quantum, or ``Mock Quantum'' Theory} 
\label{s:2}
In this section, we summarize the construction of Ref.~\cite{Minic:2014zsa}
that was motivated by certain fluid dynamics experiments critically reviewed in \cite{droplet, fluidpilot}.
The major point of the following presentation is to demonstrate that a Hamiltonian classical system can be written
in a way that resembles the formulation of quantum theory, through the usage of wavefunction-like
variables defined on the phase-space of the considered classical system.
The crucial difference between this formulation of classical theory and the canonical Schr\"{o}dinger equation of quantum theory,
is the appearance of certain non-linear and non-local terms. This motivates our proposal or, inspiring hypothesis for an emergent (mock, or analog) quantum theory
in complex adaptive systems that might be advantageous for such systems from an evolutionary point of view, 
by reconciling adaptability with stability in an optimal way, 
in the presence of a system-dependent environment that cancels the said non-linear and non-local terms~\cite{footnote1}
and implies the emergence of an effective, linear Schr\"{o}dinger equation.

One way to visualize this is that in, say, a cellular environment, there exist oscillatory wave processes which get tuned in to
the classical particle-like dynamics of the cellular matter, so that the effective description is a ``wave-particle'' collective entity
the dynamics of which is described by an emergent Schr\"{o}dinger equation, with an emergent Planck constant. To describe this more precisely, in equations, we start with the Hamilton-Jacobi picture of classical dynamics, 
\be
\frac{\partial S}{\partial t}
+H\Big(Q_i,\, P_i \define \frac{ \partial S}{\partial Q_i}\Big)=0. 
 \label{dS+H=0}
\ee
The action $S$  is defined by the canonical expression ($Q_i$ denotes the configuration variables and $P_i$ their conjugate momenta)
\be
S= \int (P_i \dot{Q}_i - H) \rd t.
\ee
According to the Liouville theorem, for such a closed system the phase space volume is conserved (the volume density of phase space being $\rho$)
\be
\frac{\partial \rho}{\partial t} + \nabla(\rho \vec{v}) =0,
 \label{Liouville}
\ee
where the velocity $\vec{v}$ is defined as $v_i \define \dot{Q}_i$, and in the single particle case $\vec{v} = \frac{\nabla S}{m}\define \frac{\vec{P}}{m}$.
Following \cite{rs} we define the 
new variables
\be
\psi \define \pm \sqrt{\rho} \,\exp\Big(i \frac{S}{\mockbar}\Big).
\label{psi}
\ee 
The mock (or effective) Planck constant $\mockbar$ appears by dimensional arguments given this definition and the application-specific
dimension (physical units) of the action $S$.  (For example, in the case of the Lotka-Volterra dynamics the relevant 
action is dimensionless \cite{lv, other}, and thus analogous to the eikonal of wave optics.)
The variable $\psi$ in~\eqref{psi} satisfies the following ``$\psi$-equation'' \cite{rs} 
\be i \mockbar
\frac{\partial \psi}{\partial t}
  = H \Big(Q_i,\, \frac\mockbar{i} \frac{\partial}{\partial Q_i}\Big) \psi 
  - V_Q (\psi, \psi^*)\, \psi.
\label{mq1}
\ee
There is also the complex conjugate equation for $\psi^*$.  Note that in this expression $V_Q$ is the so-called
``quantum'' potential of de-Broglie and Bohm \cite{holland, bohm, rpf}, which depends on $\psi$ and $\psi^*$ and
stems from the kinetic term in the action. In the case of the canonical (single particle) kinetic term
$\frac{P_i^2}{2m}$, the quantum potential is 
\be
V_Q = -\frac{{\mockbar}^2}{2m} \frac{\nabla^2
  (\sqrt{\rho})}{\sqrt{\rho}}.
 \label{VQ}
\ee
We will discuss a more general form of the quantum potential later in this paper.
The above non-local and non-linear dynamics generally evolves pure states into mixed states, and so is not coherent (and not unitary) in the sense of the canonical linear ``quantum'' theory \cite{rs}.

For a realistic complex system, such deterministic dynamics should be supplemented with an external source of noise or other
environmental factors not included in the model, ${\eta}$, in which case the stability analysis is generalized from
the deterministic to a stochastic analysis \cite{stability}. 
(For concepts of metastability and multi-stability that enrich the dynamics of complex systems see \cite{kelso2012}. The control of multi-stable systems is of great practical importance and it is an active area of research \cite{pisarchik}.)

The mock quantum theory proposal then posits that it is advantageous for an adaptive complex system to develop this new type of
``quantum'' stability and linearity and show how it can emerge if the environmental/stochastic source $\eta(\psi, \psi^*)$ cancels against
the non-linear and non-local part of the ``$\psi$-equation,'' 
turning it thereby into an emergent and effective
Schr\"{o}dinger equation. In the presence of such external/environmental noise term, $\eta(\psi, \psi^*)$, the equation~\eqref{mq1} should be modified to
\be i \mockbar \frac{\partial \psi}{\partial
  t} = H \Big(Q_i, -i \mockbar \frac{\partial}{\partial Q_i}\Big) \psi
    \underbrace{ - V_Q (\psi, \psi^*)\, \psi
                  +\eta(\psi, \psi^*) }_{\approx 0},
\label{cancel}
\ee 
where we indicate that the $V_Q (\psi, \psi^*) \psi$ and ${\eta(\psi, \psi^*)}$ can be combined into one effective term
which according to our proposal cancels in a complex adaptive environment. 
Since such cancellation would be
difficult to achieve if $\eta(\psi, \psi^*)$ were purely stochastic, 
one should expect the environment also to be adaptive. 
(As $\eta(\psi, \psi^*)$ is expected to contain a stochastic component, we will return to the consequences of imperfect cancellation in~\eqref{cancel}.) 
This is somewhat reminiscent of
the apparently observed de-Broglie-Bohm-like behavior of droplet ``particles'' guided atop a vibrating ``pilot-wave'' ripple-tank 
of a classical fluid \cite{droplet, fluidpilot} --- this experiment is described in more detail in Appendix~\ref{a:Hydro}.
The process by which
mock-quantum framework emerges is in essence the reverse of the decoherence approach to the quantum-to-classical transition \cite{zurek},
as the system as a whole ``re-coheres'' in the presence of
a ``fine-tuned" $\eta(\psi, \psi^*)$ to give a steady state non-equilibrium dynamics captured by an effective Schr\"{o}dinger equation and the associated Born rule for probabilities. 
(Note that the quantum potential is subtracted by the adaptive environment in the context of $\psi$ variables, but that it is effectively added to the original Hamiltonian in terms of the original classical variables.)

The above construction evidently relies on a ``balancing act,''
which might be a generic problem of mock quantum theory,
perhaps analogous to the generic problem of quantum theory
in macroscopic systems, to wit, the problem of decoherence.
However, the cancellation between the quantum potential term and the environment does not have to be perfect, and realistically never is.
Such a noise term can be taken into account by using stochastic equations. 
We will comment on the formalism that takes into account noise terms in Section~\ref{s:7} of this paper. 
Until then we proceed by assuming perfect cancellation in order to illustrate one of our main points, the state-dependent dynamics of mock quantum theory. Nevertheless, one might
ask what happens to the emergent Schr\"{o}dinger equation under imperfect cancellation in 
Eq.~\eqref{cancel}, i.e., by adding a perturbation to the mock Schr\"{o}dinger equation.
The natural proposal here is that precisely such perturbations will lead to the ``collapse''
of the emergent ``wavefunction'' and the actually observed values of measured quantities,
with the probability distribution governed by the Born rule.
Perturbations in the (adaptive) noise would thereby
be crucial for a dynamical ``collapse'' of the emergent Schr\"{o}dinger ``wavefunction'' along the lines
of various proposals reviewed in \cite{bassi}. 
(The crucial role of noise in the emergence of classical behavior in the
de-Broglie-Bohm interpretation of the canonical quantum theory is nicely summarized in \cite{bohm}.)
We will discuss the generic stochastic mock quantum dynamics in Section~\ref{s:7}. In that section we will show that stochastic mock-quantum dynamics is equivalent to a state-dependent statistical field theory.

A fundamental reason for the advantageous nature of
emergent ``quantum'' theory might be found in the linear structure of quantum theory, which is tightly related to
the concept of of maximally symmetric Fisher information \cite{wootters,chia}.  In an evolving environment the adaptive dynamics of complex
systems would tend to adjust the whole system
(for competing reasons of adaptability and stability) so that an effective or mock-quantum theory emerges as
\be i
\mockbar \frac{\partial \psi}{\partial t} 
 = H \Big(Q_i,\, \frac\mockbar{i} \frac{\partial}{\partial Q_i}\Big) \psi.
 \label{SchEq}
\ee
(This emergent linear quantum evolution could be also motivated by the cellular automaton interpretation of quantum theory developed in \cite{thooft}; see also: \cite{othereq,thooft1,cellular}.)
 Such mock quantum theory would be probabilistic and it would map the mock wave function
 to the observed phenomena via the mock Born rule $|\psi|^2$.
This expression should be understood as a steady state distribution of states for the non-equilibrium dynamics of the underlying classical system. This picture is also consistent with the experiments (whose validity is still critically discussed) on the emergent de-Broglie-Bohm dynamics in fluid dynamics \cite{droplet, fluidpilot}, (see Appendix~\ref{a:Hydro}). Such mock Born rule would allow for the crucial {\em interference} of probabilities --- one of the hallmark signatures of mock-quantum behavior in complex adaptive systems --- and it might be the underlying
reason for the recently observed usefulness of quantum probabilities in various complex classical systems \cite{plotnitsky,khrenikov}.

Let us close with a brief comment on the cancellation condition~\eqref{cancel}. Combined with~\eqref{VQ}, it implies
\begin{equation}
   \frac{{\mockbar}^2}{2m}(\nabla^2\sqrt{\rho}) \,e^{iS/\mockbar} \approx \pm\,\eta(\psi, \psi^*),
    \quad\text{i.e.,}\quad
   \bigg\{\begin{array}{@{}r@{~}l}
     (\nabla^2\sqrt{\rho}) &\approx 2m\,|\eta|\,/{\mockbar}^2,\\[4pt]
  (S/\mockbar)
    &\approx \big(2n{+}\genfrac\{\}{0pt}{1}10\big)\pi
        +\tfrac1{2i}\log(\eta/\eta^*),
          \end{array}
 \label{rho-S}
\end{equation}
which provides external/environment noise $\eta$-driven constraints on Hamilton's action, $S$, and the phase space volume density, $\rho$ --- in addition to the dynamical equations~\eqref{dS+H=0} and~\eqref{Liouville}, i.e., in addition to~\eqref{mq1}.
 A plausible interpretation of this is that the external/environment noise, $\eta$, {\em\/effectively selects,} via this additional constraint, from among the (classically continuous) distribution of possible states of the $(Q_i,P_i)$-system:
 At such $\eta$-determined values of the action and phase-space volume-density, $(S,\rho)$, the dynamics of the $(Q_i,P_i)$-system simplifies significantly: Eq.~\eqref{cancel}$\,\to\,$\eqref{SchEq}.
 In the following two sections we will present a more generic argument as to why the non-equilibrium dynamics of a classical system could lead to the above mock quantum description.

\section{Quantum, Mock Quantum and a Comparison with de-Broglie-Bohm}
\label{s:3}

Now we are ready to address the elements of the mock quantum proposal/hypothesis in more detail.
In Appendix~\ref{a:B} we review some basic facts about different formulations of quantum and mock quantum theory and in this section we concentrate on the mock de-Broglie-Bohm formulation and the concept of the mock quantum potential, which is central for the rest of the paper.

In order to make a connection between the mock quantum description and the underlying classical dynamics let us rewrite the (mock) Schr\"{o}dinger equation in the polar decomposition (this is also the polar decomposition of the
path integral), as this rewriting will turn out to 
be particularly useful in the rest of the paper
\cite{holland, bohm}
\be
\psi \define \pm \sqrt{\rho} \:e^{iS/\mockbar},
\ee 
whereby the complex Schr\"{o}dinger equation becomes a coupled system of two real equations, already given in the previous section.
The first one is the equation of continuity,
\be
\frac{\partial \rho}{\partial t} + \nabla{\cdot}(\rho \vec{v}) =0,
\label{dBB.rho}
\ee
where the velocity $\vec{v}$ is defined as $v_i \define \dot{Q}_i$, and in the single particle case $\vec{v} = \frac{\nabla S}{m}\define \frac{\vec{P}}{m}$.
The second equation is the de-Broglie-Bohm correction to the classical Hamilton-Jacobi equation
\be
\frac{ \partial S}{ \partial t} + H 
+ V_Q =0,
\label{dBB.S}
\ee
where $V_Q$ is the quantum potential.
Whenever $H = \frac{p^2}{2m} + V(q)= \frac{{\nabla S}^2}{2m} + V(q)$, 
then $V_Q = -\frac{{\mockbar}^2}{2m} \frac{\nabla^2
  (\sqrt{\rho})}{\sqrt{\rho}}$
is given by direct computation. Note that the quantum potential is added to the classical Hamiltonian.
These equations, supplemented by the Schr\"{o}dinger equation (because the quantum potential needs explicit wavefunctions that are
computed by solving the Schr\"{o}dinger equation) provides a non-covariant rewriting of quantum theory in a particular basis in which both
the classical (particle) and the quantum (wave) variables appear.

Perhaps the easiest (as well as more general and illuminating way) to derive these equations
is to use the density matrix formalism (following \cite{hiley}).
In particular, let us start from the Schr\"{o}dinger equation written for $\rho \define |\psi \rangle \langle \psi|$
where (with $\mockbar =1$, for simplicity)
\be
i \frac{\partial |\psi \rangle}{ \partial t} = H |\psi \rangle, \quad - i \frac{\partial \langle\psi |}{ \partial t} =  \langle \psi | H.
\ee
The difference of the $\langle\psi|$- and $|\psi\rangle$-multiples
of these two equations yields the analog of
the Liouville equation in classical theory, which can be written for the density on phase space, and similarly, in the case of quantum theory one has
for the density matrix operator $\hat{\rho}$ (with a unit mock Planck constant, and a Hamiltonian operator $\hat{H}$)
\be
i \frac{\partial \hat{\rho}}{\partial t} + [\hat{\rho}, \hat{H}] =0,
\ee
where $[\,,]$ is the usual commutator 
$[\hat{\rho}, \hat{H}] \overset{\sss\rm def}= \hat\rho \hat{H} - \hat{H} \hat\rho$.
Now, as pointed out in \cite{hiley}, the sum of the above multiples of the Schr\"{o}dinger equation and its conjugate implies:
\be
i \Big(\frac{\partial |\psi \rangle}{ \partial t}   \langle \psi | -  |\psi \rangle \frac{\partial \langle\psi |}{ \partial t} \Big) 
= \hat\rho \hat{H} + \hat{H} \hat\rho.
\ee
The same polar decomposition
\be
\hat\rho \define \hat{R}^2 , \quad 
|\psi \rangle = \hat{R}\,\hat{U}, \quad
\langle \psi | = \hat{U}^{\dagger}\,\hat{R}, \quad
\hat{U}\,\hat{U}^{\dagger} = 1,
\ee
then implies
\be
i \Big(  \frac{\partial |\psi \rangle}{ \partial t}   \langle \psi | 
        -|\psi \rangle \frac{\partial \langle\psi |}{ \partial t} \Big) 
\define 
i \hat{R} \Big(\frac{\partial\hat{U}}{ \partial t}\hat{U}^{\dagger}
              -\hat{U}\frac{\partial\hat{U}^{\dagger}}{ \partial t}\Big)
= \hat\rho\,\hat{H} + \hat{H}\, \hat\rho.
\ee
Finally, writing the unitary operator $\hat{U}$ as 
\be
\hat{U} = e^{i\hat{S}}, \quad \hat{S}^{\dagger} = \hat{S}, 
\ee
produces
\be
\hat{R}\, \frac{\partial \hat{S}}{\partial t}\,\hat{R}
+ \frac{1}{2} (\hat\rho\,\hat{H} + \hat{H}\,\hat\rho) =0,
\ee
where if we assume that $[\hat{R}, \frac{\partial \hat{S}}{\partial t}]=0$,
the above equation becomes the analog of what in classical physics is the already cited Hamilton-Jacobi equation, which can be multiplied by
the classical phase space density $\hat\rho$ so that indeed
\be
\rho \frac{\partial S}{\partial t} +\rho H=0. 
\label{HamJac}
\ee
In quantum theory, a product of two Hermitian matrices has to be symmetrized (the Weyl ordering prescription) and thus we have
\be
\hat{\rho} \frac{\partial S}{\partial t} +\frac{1}{2}(\hat{\rho} \hat{H}  + \hat{H} \hat{\rho}) =0, 
\label{HamJacOps}
\ee
where we recognize the anticommutator (or the Jordan bracket) of the density matrix and the Hamiltonian.
Note that we could have as easily obtained
\be
{\hat{\rho}}^{1/2} \frac{\partial \hat{S}}{\partial t} +\frac{1}{2}({\hat{\rho}}^{1/2} \hat{H}  + \hat{H} {\hat{\rho}}^{1/2}) =0. 
\ee
This second equation leads, for the case of the simple Hamiltonian
$H = p^2/2m +V \define -\frac{\mockbar^2}{2m} \nabla^2 +V$ to the
already quoted~\eqref{VQ} expression for the quantum potential $V_Q$ (where we have restored $\mockbar$)
\be
V_Q={\rho}^{-1/2} \Big({-}\frac{\mockbar^2}{2m} \nabla^2 ({\rho}^{1/2})\Big).
\ee

However, this shows the general expression for the quantum potential that we can use for Hamiltonians with a non-canonical kinetic term,
like the one we will encounter in the Lotka-Volterra system in the next section:
In that case, $H = a (e^Q - Q) + d (e^P -P)$, rendering the quantum potential for the Lotka-Volterra system, $V_Q^{LV}$,
be given by the following formula
\be
V_Q^{LV}={\rho}^{-1/2} \big[ d (e^P -P)  ({\rho}^{1/2})\big], \quad 
P = -i \mockbar \nabla.
\ee

The de-Broglie-Bohm system of equations~\eqref{dBB.rho}--\eqref{dBB.S} is completely equivalent to the Schr\"{o}dinger formulation~\eqref{SchEq}.
Also, note that the (mock) quantum potential is added to the classical Hamiltonian.
{Thus, in mock, or analog, or emergent, quantum theory rewritten \`{a} la de-Broglie-Bohm, one retains the usual de-Broglie-Bohm formalism but, again, replaces
the Planck constant with an emergent, or analog, non-universal and system-dependent, mock Planck constant
$\hbar \to \mockbar$.}

In the context of de-Broglie-Bohm formalism there exists a claim that the Born rule $|\psi|^2$ is
emergent from the underlying non-equilibrium dynamics, as a steady-state distribution \cite{bohm}.
This is consistent with our proposed interpretation of the emergent Born rule in the context of mock quantum theory.

In conclusion of this section,
we note that simple, classical systems follow 
``classical trajectories,'' between given initial 
and final states. Such classical trajectories obey the least action principle. In contrast, a complex adaptive system might
have many available (and energetically equivalent)
paths between given initial and final states, that are also very sensitive to the environment of the complex adaptive system under consideration.
The presence of many trajectories of a complex adaptive system might be more effectively described using quantum measures represented by an effective Schwinger variational principle, and an effective Feynman path integral, which opens the possibility for an emergent quantum description of complex adaptive systems with an emergent, system-dependent, non-universal Planck constant.
This picture is also consistent with an empirical view that data describing complex systems do not always conform with the laws of classical probability, but rather with quantum-like probability \cite{plotnitsky,khrenikov}. 
Furthermore, this viewpoint fits into a more general picture of quantum mechanics as quantum measure theory
\cite{sorkin}.

\section{Emergence of mock quantum description in classical systems}
\label{s:4}

The central question we address here is how and why would the non-equilibrium dynamics of a classical system be described by the mock quantum description.
The original argument for mock quantum theory is based on Rosen's old observation of ``reverse-engineering'' the de-Broglie-Bohm analysis,
that is, in rewriting classical Liouville and Hamilton-Jacobi equations
in terms of ``wave functions''. In that case (for the canonical Hamiltonian)
the quantum potential emerges as a ``thing to be canceled by an adaptive
(wave function-dependent) `environmental' term.'' But why should that happen and why would the non-equilibrium dynamics of a classical (complex and adaptive) system lead to a mock quantum description? The only existing empirical evidence for this possibility at the moment is the fluid mechanics experiment described in~\cite{droplet, fluidpilot} (see Appendix~\ref{a:Hydro}), for which, if correct, no purely classical approach predicts the behavior of the system. In that context, a droplet is carried by a wave (thus simulating the ``pilot wave'' picture of de Broglie~\cite{holland}) and
the claim of the original experiment is that~\cite{droplet, fluidpilot}, such a system exhibits emergent quantum properties, as encountered in the double-slit interference experiment. 
The original argument for mock quantum theory, motivated by this experiment~\cite{droplet, fluidpilot} and Appendix A, is based
on Rosen's old observation of ``reverse-engineering'' the de-Broglie--Bohm analysis,
that is, in rewriting classical Liouville and Hamilton-Jacobi equations
in terms of ``wave functions''. In that case (for the canonical Hamiltonian)
the quantum potential emerges as the departure from the quantum behavior and hence becomes the``thing to be canceled by an adaptive
(wave function-dependent) `environmental' term in order to obtain the mock-quantum behavior.'' But why should that happen? 

To answer this question, it is not enough to look at the Liouville and the Hamilton-Jacobi equations by themselves. 
As emphasized in the part of Section~\ref{s:3} that summarizes the operatorial argument for the quantum potential, one needs to examine the averaging of the Hamilton-Jacobi equation~\eqref{HamJac} by the probability density that satisfies the Liouville equation.
To that end, the product of classical quantities $\rho$ and $H$ should be replaced by the symmetrized product (the anticommutator) of the corresponding operators~\eqref{HamJacOps},
which then produces the quantum potential.

Let us examine this last assertion more closely:
For a generic non-equilibrium system, with dissipation (that is, an
open system, in an environment) there is the celebrated BRS (Bowen-Ruelle-Sinai or Sinai-Ruelle-Bowen) measure 
construction~\cite{SRB=BRS}, 
as summarized by David Ruelle~\cite{Ruelle:2004con} .
The crucial point is that generically, a non-equilibrium system in a
dissipative environment will evolve into an attractor (again, generically, a
fractal) in its phase space and that on that attractor one will have a (BRS)
probability measure that satisfies the ergodic theorem: that is, the time
averages will be equal to the averages over that probability measure $\rho$, that is properly normalized $\int \rho =1$
\be
\lim_{T \to \infty} \frac{1}{T} \int \rd t\; f(q(t), p(t), t) 
= \int \rho\; f (q, p).
\ee
Thus because the attractor is a fractal, the product
between $\rho$ and $H$ is not the usual product of functions, but, in general,
a deformed product on that fractal phase space. This point was not elaborated on in our original work \cite{Minic:2014zsa}.

Next one should ask, what deformed product obeys the symmetries (symplectic
transformations) of the classical phase space. It is well known that the
Poisson brackets (the generators of symplectic transformations which
are responsible for the Liouville theorem in the first place) are uniquely
deformed into Moyal-brackets (essentially, exponents of Poisson brackets, with an
effective mock Planck constant), which in turn can be mapped onto commutators of operators
on a Hilbert space(for a nice review  of this topic see 
~\cite{Curtright:2011vw}.)
Therefore the product of $\rho$ and $H$ on a fractal attractor in the phase
space of a generic non-equilibrium system would be mapped into
a symmetric version of Moyal bracket, that is the anti-commutator
of operators $\rho$ and $H$, and thus, equation (21) of the previous section.
In other words, if we use~\cite{Curtright:2011vw}
\begin{equation}
  a\star b \define a\big(q,p-\tfrac12i\mockbar\overset{\rightarrow}{\partial_q}\big)\,
   b\big(q,p+\tfrac12i\mockbar\overset{\leftarrow}{\partial_q}\big),\qquad
  [a,b]_\star \define a\star b - b\star a,
 \label{star}
\end{equation}
to denote the deformed product of $ab$ and their Moyal bracket, we have
\be
\rho H \equiv \frac{1}{2} (\rho H + H \rho) \to \frac{1}{2} (\rho\star H + H\star \rho) \to \frac{1}{2} (\hat{\rho} \hat{H} + \hat{H} \hat{\rho}).
\ee
This sequence leads from the averaged Hamilton-Jacobi equation (20) to the operatorial equation (21), which implies the
mock quantum potential, which, in turn, would for a simple canonical Hamiltonian generate the
canonical quantum potential!
This is a kind of  ``averaged Rosen argument'' that uses 
generic properties of fractal attractors on the phase space of 
non-equilibrium systems that lead to ergodicity.

The argument above is the real rationale for mock-quantum theory, and the answer to the crucial question:``why would a classical non-equilibrium dynamics lead to a mock quantum description?''
Furthermore, this also implies the emergence of the Born probability
measure $\rho = |\psi|^2$, which is manifestly positive definite, as a stationary non-equilibrium measure.

Now if we revisit the ``droplet-carried-by-a-wave'' experiment in the fluid mechanics, which has been repeated several times with different results \cite{droplet, fluidpilot}, we can claim that if we choose a suitable Hamiltonian $H$ for this system, we
would get the general reason why mock quantum description should
emerge in that experiment, in the limit in which the ergodic theorem is satisfied.
Of course, as BRS say, there are many non-Gibbsian measures, and
ours is just one that is consistent with the properties of the geometry
of the non-equilibrium, fractal, phase space. So there should exist
a ``mock-quantum-universality-class'' for certain non-equilibrium systems
(such as the one represented by the experiment~\cite{droplet, fluidpilot})
in particular environments that support the above emergence scenario
for the effective mock quantum potential.
It is possible that ``a fine tuning of the environment'' is needed for
such a universality class. However, the mock quantum description preserves the individual properties of
the classical system (as indicated by the averaging of the Hamilton-Jacobi equations). Also the emergent mock quantum
probability is given by the Born rule and the emergent quantum like properties. This is a distinguishing and robust feature of
mock quantum theory as opposed any other classical BRS probability measure of a generic classical non-equilibrium system.

Note that in the above argument, $H$ is the Hamiltonian of the original system. Fractality comes from the competition of chaos (generic nonlinearity) and dissipation 
provided by the interaction of the open system and the environment. We take stochasticity into account later in the paper when we discuss
mock quantum stochastic field theory (Section~\ref{s:7}). 

The real reason for mock quantum emergence is when a complex adaptive system becomes ergodic and that could be the case say for natural
selection/evolution.
So the real difficulty in that fluid experiment \cite{droplet, fluidpilot} with ``a droplet 
riding a wave'' type of setup is in ergodicity.
Are the time averages equal to averages over
some distribution? If ergodicity is achieved in
the experiment (which might require long times),
then the stationary measure is the Born measure
of mock quantum theory.

So, suppose one performs an experiment
like the fluid experiment with that droplet riding a wave, or
a biological experiment that we think might lead to a quantum
like behavior. If the time average of gathered data can be expressed as
the ensemble average of the same data (with the ensemble average
being given by the Born rule) then we can say that the system is
ergodic in the sense of mock quantum theory. This would be 
an empirical criterion for mock quantum theory. 
The crucial thing here is that in the ergodic limit
the average of the equation of motion should be such that
the properties of the equation of motion are maintained
(that is why the Poisson bracket is deformed into the Moyal bracket etc).
Thus individual stability is maintained even as the system adapts to its environment.
This mechanism should be advantageous from the point of view of natural selection which should favor complex systems that are stable but adaptable to their environment.

To conclude, the essential new insight 
is as follows: Unlike the usual statistical averaging (for example, by using the equilibrium Gibbs distribution or any other equilibrium
or non-equilibrium distribution), mock quantum averaging in the sense
of Born maintains the properties of individual systems that are being
averaged over. This is simply not true in canonical equilibrium statistical physics where we
forget about individual systems by the procedure of coarse-graining. 
That is the crucial property that would realize Schrodinger's ``order from disorder'' prescription~\cite{es},
that might be needed for the non-equilibrium physics of living systems. For example, metabolic reactions have complex pathways, but they are all individually stable even though
they happen in an environment of so many other metabolic reactions which
in principle could destroy that individual stability ~\cite{metabolism}. What mock quantum
could do for biological systems is to 
provide a far-from-equilibrium distribution that is 
consistent with ergodicity while maintaining the individual properties
of averaged processes (such as individual stability of those processes).
This is something special and that is why the averaged equations of motion 
get deformed, in the ergodic limit, into the mock quantum equations of motion, 
containing the quantum potential, while producing a probability distribution
(given by the Born rule of mock quantum theory) needed for that averaging in the sense of stationary non-equilibrium dynamics.
However, a more detailed discussion of specific biological systems (such as metabolic reactions) is beyond the scope of this paper. 
Nevertheless, we could say that
evolutionary dynamics and natural selection might exploit mock quantum description for the reasons of stability that is compatible with adaptability.

\section{The Lotka-Volterra system}
\label{s:5}
In this section we discuss our prime example: the Lotka-Volterra model.
We concentrate on this example because it is a simple biological model with a known Hamiltonian form, and it is also related to other prominent models such as Wilson-Cowan model in neuroscience~\cite{WC}.
The deterministic Lotka-Volterra system \cite{lv} specifies the number of species and how their populations interact and change in time. For the case of two species with populations $N_i (t), i=1,2$ the Lotka-Volterra equations read \cite{lv, other}
\be
\dot{N_1} = a\,N_1 - b\,N_1N_2 ~~,~~~~ 
\dot{N_2} = c\,N_1N_2 - d\,N_2  ,
\label{LV}
\ee
where $\dot{N_i}$ denotes the time ($t$) derivative, $a$ and $d$ are the relevant auto-increase or auto-decrease parameters, $b$ and $c$ denote the interaction strength between the species ($a,b,c,d>0$)
 \cite{lv, other}.
The stationary ($\dot{N_i}=0$)  population levels $N_i = q_i$, apart from the trivial (0,0), occur for $q_1=d/c $ and $q_2=a/b$.
It is known that the Lotka-Volterra system, for small initial displacements from the equilibrium point ($q_1, q_2$), will oscillate with the frequency 
\be
\omega=\sqrt{ad}.
\label{freq}
\ee
The Lotka-Volterra system (\ref{LV}) has a Hamiltonian, which can be written as follows \cite{other}: First we introduce dimensionless
$
z_i \define \log( N_i/q_i) .
$
Then 
equations (\ref{LV}) read as:
\be
\dot{z}_1= a(1-e^{z_2})~~,~~~ \dot{z}_2= d(e^{z_1} -1).
\label{LV2}
\ee
A time-invariant function $H$ for the system (\ref{LV2}), which has the property that 
\be
\dot{z}_1= -\frac{\partial H}{\partial z_2}~~,~~~ \dot{z}_2= \frac{\partial H}{\partial z_1},
\ee
is the Lotka-Volterra Hamiltonian \cite{other}:
\be
H = a (e^{z_2} -z_2) + d (e^{z_1} - z_1) .
\label{hlv0}
\ee
Identifying the canonical pair of variables as
$z_1= P$, and $z_2=Q$,
rewrites the Lotka-Volterra Hamiltonian \cite{other} 
\be
H =  a (e^Q -Q) + d (e^P - P).
\label{hlv}
\ee
The deterministic Lotka-Volterra equations are thereby rewritten as the canonical Hamilton equations:
\be
\dot{Q} = \frac{\partial H}{\partial P}, \quad 
\dot{P}= -\frac{\partial H}{\partial Q},
\ee

In the stationary case, the emergent quantum theory
of Lotka-Volterra dynamics advocated in this work is characterized by the emergent stationary Schr\"{o}dinger
equation
\be
H \psi_n = E_n \psi_n, \quad 
\psi_n(t) = \exp\Big({-}\frac{i}{\mockbar} E_n t\Big)\,\psi_n(0),
\ee
 where $E_n$ are the emergent eigenvalues of $H$ and where $n=0,1,2...$.
Now, treating $P$ and $Q$ as small yields the linearized approximation of the
slow Lotka-Volterra dynamics, $e^Q \sim 1+ Q +\tfrac12Q^2$ and similarly 
$e^P \sim 1+ P +\tfrac12P^2$, whereby the effective
Lotka-Volterra dynamics becomes harmonic \cite{other}:
\be 
H \sim (a+d) + \tfrac{1}{2} (a Q^2 + d P^2 ).  
\ee 
The mock quantum theory is then mapped to the simple quantum-like
harmonic oscillator with energy levels given by the standard result for the ``energy spectrum''
(as we will discuss in the next section)
\be E_n = (a+d) + \mockbar \sqrt{ad} (n +\tfrac{1}{2}) 
= (a+d) + \mockbar \omega (n +\tfrac{1}{2})~, 
\ee 
where $\mockbar $ is 
the unit of the
Lotka-Volterra action.  
(The ``energy'' $E$, as well as the Hamiltonian $H$, have dimensions of the inverse of time, like the coefficients $a$ and $d$, whereas $Q$ and $P$ are dimensionless, implying that {$\mockbar$ is dimensionless, as a relative coefficient between commensurate constants: $[\mockbar]=\frac{[a{+}d]}{[\sqrt{ad}]}$.})
The corresponding ``stationary wavefunctions'' $\psi_n$ are given in
terms of the appropriate Hermite polynomials.  In this case all transition probabilities can be computed exactly.
In principle, the classical variables $Q$ and $P$ become the canonically normalized operators $\hat{Q}$ and $\hat{P}$, obeying the canonical commutation relations $[\hat{Q}, \hat{P}] = i \mockbar$. For a given ``quantum'' state, this commutation relation in turn implies the canonical indeterminacy relation between $\Delta Q$ and $\Delta P$. Therefore in the emergent ``quantum'' phase the Lotka-Volterra variables $Q$ and $P$ and therefore $z_1$ and $z_2$, cannot be simultaneously measured with absolute precision while preserving the state of the system. However, we emphasize that $\mockbar$ is 
system dependent and non-universal, and thus this statement of the ``mock indeterminacy principle'' is also system-dependent and
non-universal.

In the case of the above  harmonic oscillator (H.O.) approximation, for the case of
the stationary states with the energy $E_n$, the quantum potential is simply given by \cite{holland} 
\be
V_{Q}^{(H.O.)} = \mockbar \omega (n +\tfrac{1}{2}) - \tfrac{1}{2} \omega Q^2.
\ee
This follows from the exact solution of the ``mock Schr\"{o}dinger equation,'' which
involves the standard Hermite polynomials.
The time dependent form of the ``mock-quantum potential'' in our situation can be also easily inferred 
from the explicit solution of the canonical quantum harmonic oscillator \cite{holland}
\be
V_{Q}^{(H.O.)} (t) = -\frac{\omega}{2} \big(Q - A \cos(\omega t)\big)^2 +
\tfrac{1}{2} \mockbar\omega.
\label{vqharm}
\ee
This expression represents the time-dependent ``holistic'' response of the environment in
our proposal, i.e. it determines the negative of the ``environmental'' 
term $\eta$.
(Recall that $\psi\,V_{Q}^{(H.O.)} (t) + \eta \to 0$, is a necessary criterion for the emergence of effective mock quantum theory,
in this case, in the harmonic oscillator approximation of the deterministic Lotka-Volterra dynamics in a suitable adaptive environment that allows for an emergent mock quantum description.)

Note that for the general Hamiltonian we have
\be
\Big[a(e^Q - Q) + d (e^{\hat P} -\hat{P})\Big] \psi_n = E_n \psi_n,  \quad 
\hat{P} = -i \mockbar \nabla.
\ee
In this stationary Schr\"{o}dinger equation the position $Q$ and the momentum $P$ enter on the same footing, so the
equation should be factorizable, at least for the ground state.
We will comment on this in the following section.

Alternatively, we could write the de-Broglie-Bohm dynamics for this ``mock qantum'' Lotka-Volterra system and examine 
its behavior (by using, in general, numerical simulations, because of the non-local and ``holistic'' nature of the ``quantum" potential.) That is the subject of the next section of this paper.

\section{The Mock Quantum Potential and the Lotka-Volterra System}
\label{s:6}
In this section we apply de-Broglie-Bohm formalism to the mock quantum (MQ) Lotka-Volterra (LV) system for two species in order to illustrate the state-dependent nature of the mock quantum dynamics.
First, we note that according to Section~\ref{s:3} in which we have summarized the de-Broglie-Bohm formalism the classical Hamiltonian is shifted by the ``mock quantum" potential $V_Q$ which depends on the form of the Hamiltonian
as well as the eigenfunctions of the Hamiltonian, but it is essentially controlled by the momentum (kinetic, or inertial) part of the Hamiltonian.
Thus, our proposal or hypothesis regarding the emergence of a mock quantum description in the non-equilibrium dynamics of classical, complex and adaptive systems could be also summarized as follows:
the criterion of stability that is compatible with adaptability in the mock quantum limit force us
 to model the
environment’s effect as an additive perturbation
(given by the 
mock quantum potential) of the classical dynamical equations.

\subsection{Quadratic approximation}

Let us start with the quadratic approximation.
In that case the LV Hamiltonian is quadratic
\be
H_2 = \tfrac{1}{2} \big(a Q^2 + d P^2\big)
= \tfrac{1}{2} \big(a z_2^2 + d z_1^2\big) 
= \tfrac{1}{2} \Big(a \big( \log( N_2/q_2) \big)^2 
                  + d \big( \log( N_1/q_1) \big)^2 \Big).
\ee
The vacuum state of this quadratic Hamiltonian is a Gaussian
\be
H_2 \psi_0 = 0 \to \psi_0= A \exp(-B Q^2),
\ee
where the constants are fixed by the (square) normalizability of the mock vacuum wavefunction $\psi_0$ as usual.
The proper normalization here is:
$\psi_0 = A \exp(-y^2/2)$,
with $A^4 = m \omega/(\pi \mockbar)$ and $y =  x \sqrt{m \omega/\mockbar}$ for
the linear harmonic oscillator (LHO) 
Hamiltonian $H= p^2/2m +m \omega^2 x^2/2$.
Thus the quantum potential  (described in Section~\ref{s:3} for a Hamiltonian with a canonical $\tfrac12P^2$ kinetic term)
\be
V_Q={\psi_0}^{-1} \big[g \nabla^2 ({\psi_0})\big] 
= \tfrac{1}{2} \mockbar \sqrt{a d} - \tfrac{1}{2} \sqrt{ad}\, Q^2,
\ee
is quadratic in $Q$. (Here, $g$ stands for $ -\frac{\mockbar^2}{2m}$, where $\mockbar$ is the dimensionless mock Planck constant appropriate for the LV dynamics, and $\omega=\sqrt{a d}$.) The constant term represents the ``vacuum energy.''
The de-Broglie-Bohm Hamiltonian is the sum of the original quadratic Hamiltonian $H_2$ and $V_Q$ and the Hamilton's equations are still linear
\be
\dot{Q} = \frac{\partial (H_2 +V_Q)}{\partial P} = d P, \quad \dot{P}= -\frac{\partial (H_2 + V_Q)}{\partial Q} = 
\Big({-}a + \frac{1}{2} \sqrt{ad}\Big)Q.
\ee
We just have to substitute $P = \log( N_1/q_1)$ and $Q = \log( N_2/q_2)$ to get the mock-quantum LV equations in this approximation.
Next we look at the higher order wavefunctions for the harmonic oscillator, $\psi_n$, given by Hermite polynomials.
For example, the first excited state is:
\be
\psi_1 = A_1 Q \exp(-B Q^2),
\ee
or more precisely:
$\psi_1 = A \sqrt{2} y \exp(-y^2/2)$,
again with $A^4 = m \omega/(\pi \mockbar)$ and $y =  x \sqrt{m \omega/\mockbar}$.
Then we have the following quantum potential that comes from the first excited state in the harmonic approximation
\be
V_Q^{(1)}={\psi_1}^{-1} \big[p \nabla^2 ({\psi_1})\big] 
= \tfrac{3}{2} \mockbar \sqrt{a d} - \tfrac{1}{2} \sqrt{ad}\, Q^2,
 \label{e:VQ1}
\ee
as expected, given the discussion from the previous section.
Similarly we have for the second excited state
\be
\psi_2 = A_2 (C Q^2 -1) \exp(-B Q^2),
\ee
or more concretely, $\psi_2= A \frac{1}{\sqrt{2}} (2y^2-1) \exp(-y^2/2)$,
again with $A^4 = m \omega/(\pi \mockbar)$ and $y =  x \sqrt{m \omega/\mockbar}$.
Formally, we have the following quantum potential for the second excited state
\be
\begin{aligned}
V_Q^{(2)}={\psi_2}^{-1} [p \nabla^2 ({\psi_2})]
 &= \frac{5}{2} \mockbar \sqrt{a d} - \frac{1}{2} \sqrt{ad} Q^2, 
\end{aligned}
 \label{e:VQ2}
\ee
where $Q = \log( N_2/q_2)$, so that we get a quadratic expression, once again.

\begin{figure}[htb]
 \begin{center}\unitlength=1mm
  \begin{picture}(175,35)(0,2)
   \put(  0, 0){\includegraphics[width=55mm]{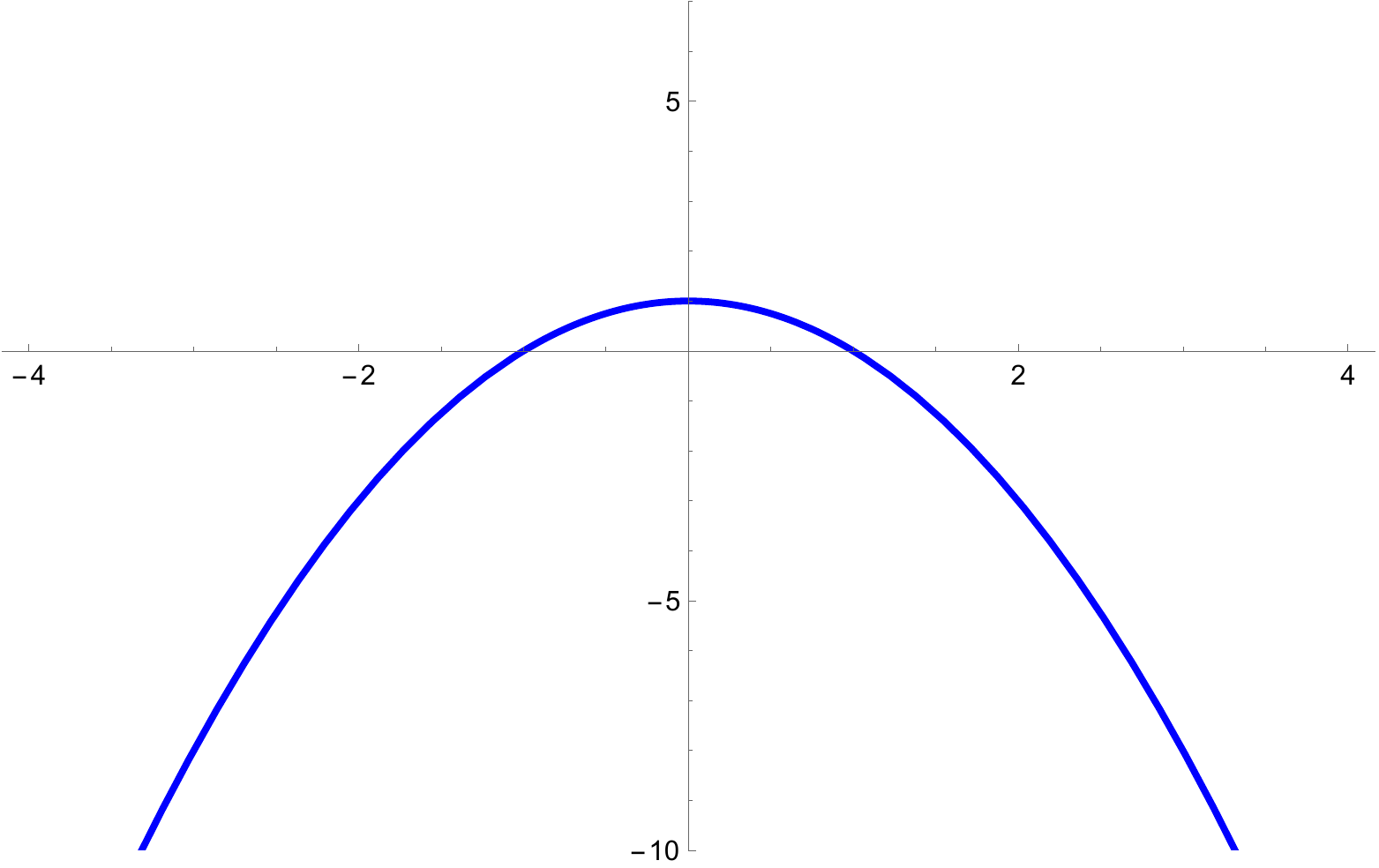}}
    \put(2,22){$n=0$}
    \put(15,5){$V_Q^{(0)}=1-y^2$}
   \put( 60, 0){\includegraphics[width=55mm]{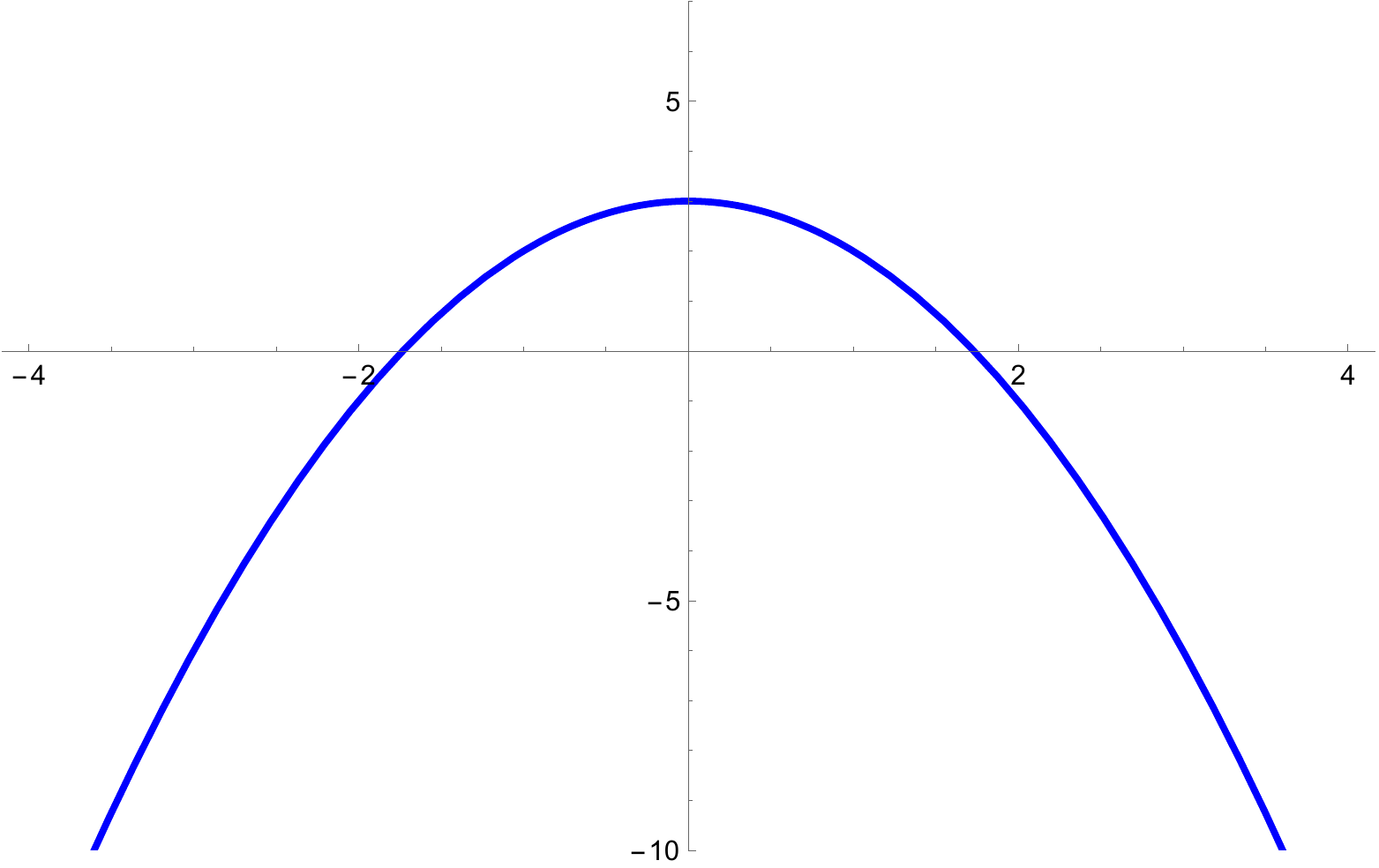}}
    \put(62,22){$n=1$}
    \put(75,5){$V_Q^{(1)}=3-y^2$}
   \put(120, 0){\includegraphics[width=55mm]{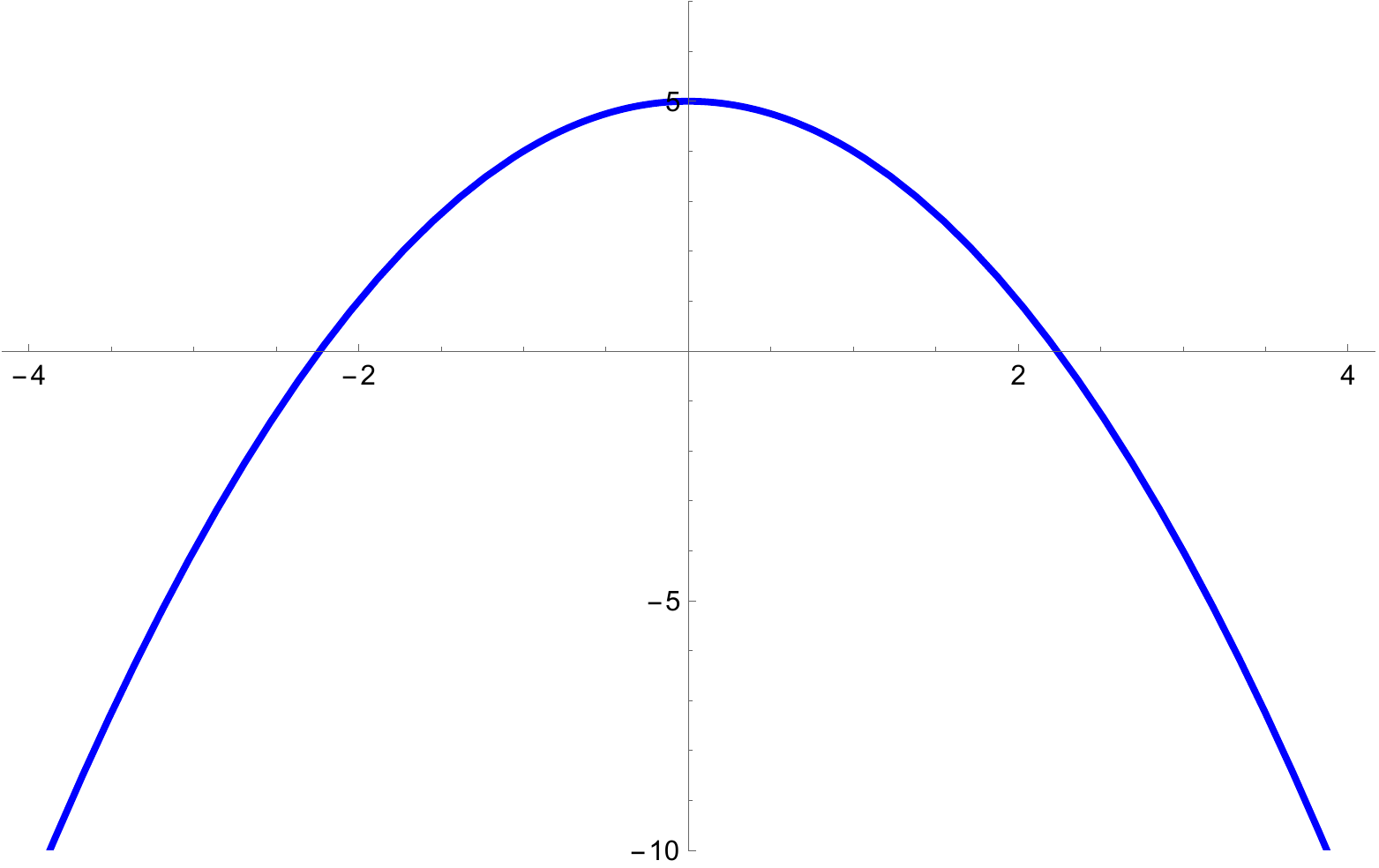}}
    \put(122,22){$n=2$}
    \put(135,5){$V_Q^{(2)}=5-y^2$}
  \end{picture}
 \end{center}
 \caption{The ``quantum potential,'' $V_Q(\psi_n)$ as defined in~\eqref{VQ}, plotted for a few lowest-energy states}
 \label{f:VQ-Hermite}
\end{figure}

\begin{figure}[htb]
 \begin{center}\unitlength=1mm
  \begin{picture}(175,35)(0,2)
   \put(  0, 0){\includegraphics[width=55mm]{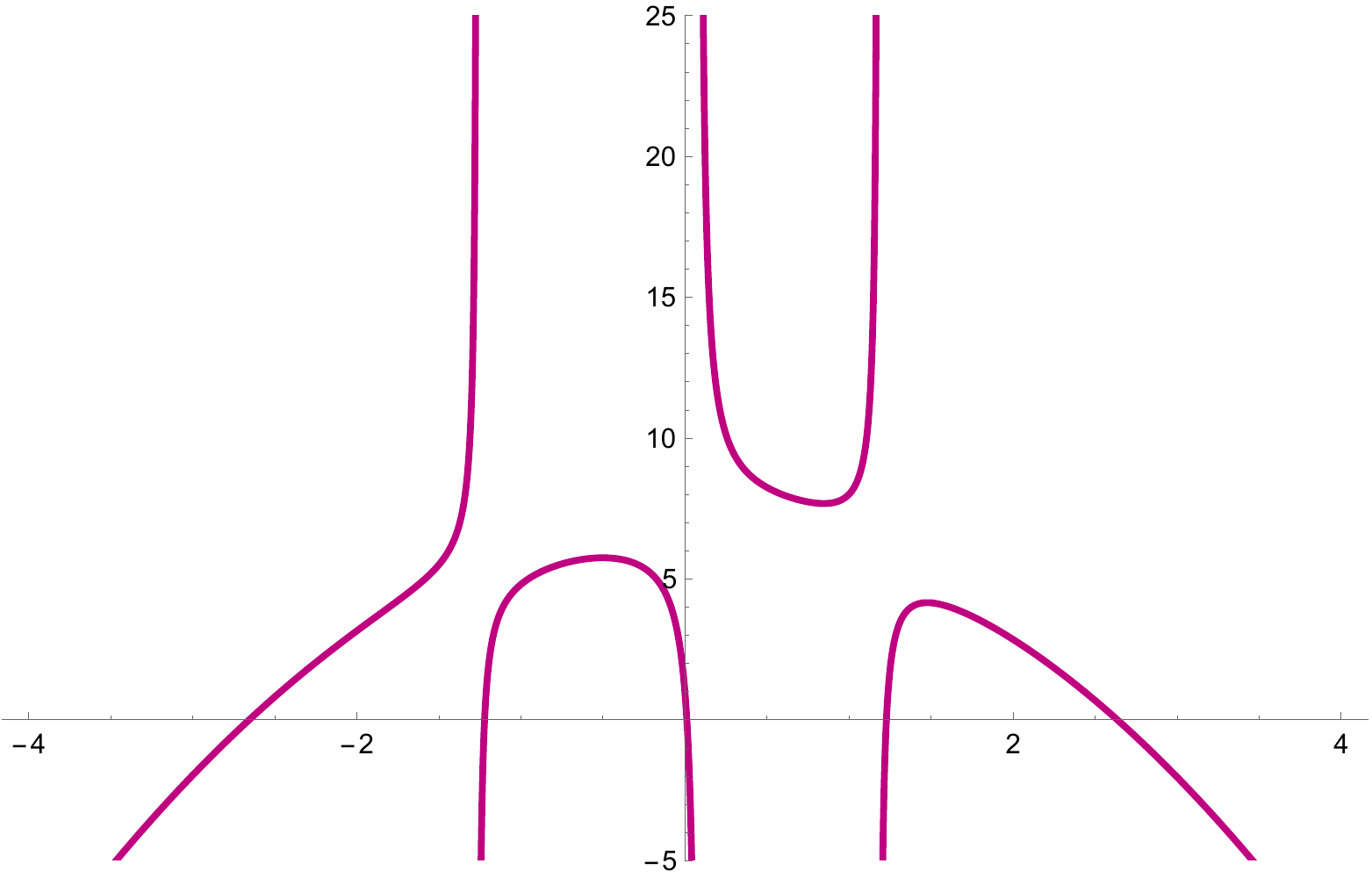}}
    \put(10,30){$V_Q^{(0,3)}$}
   \put( 60, 0){\includegraphics[width=55mm]{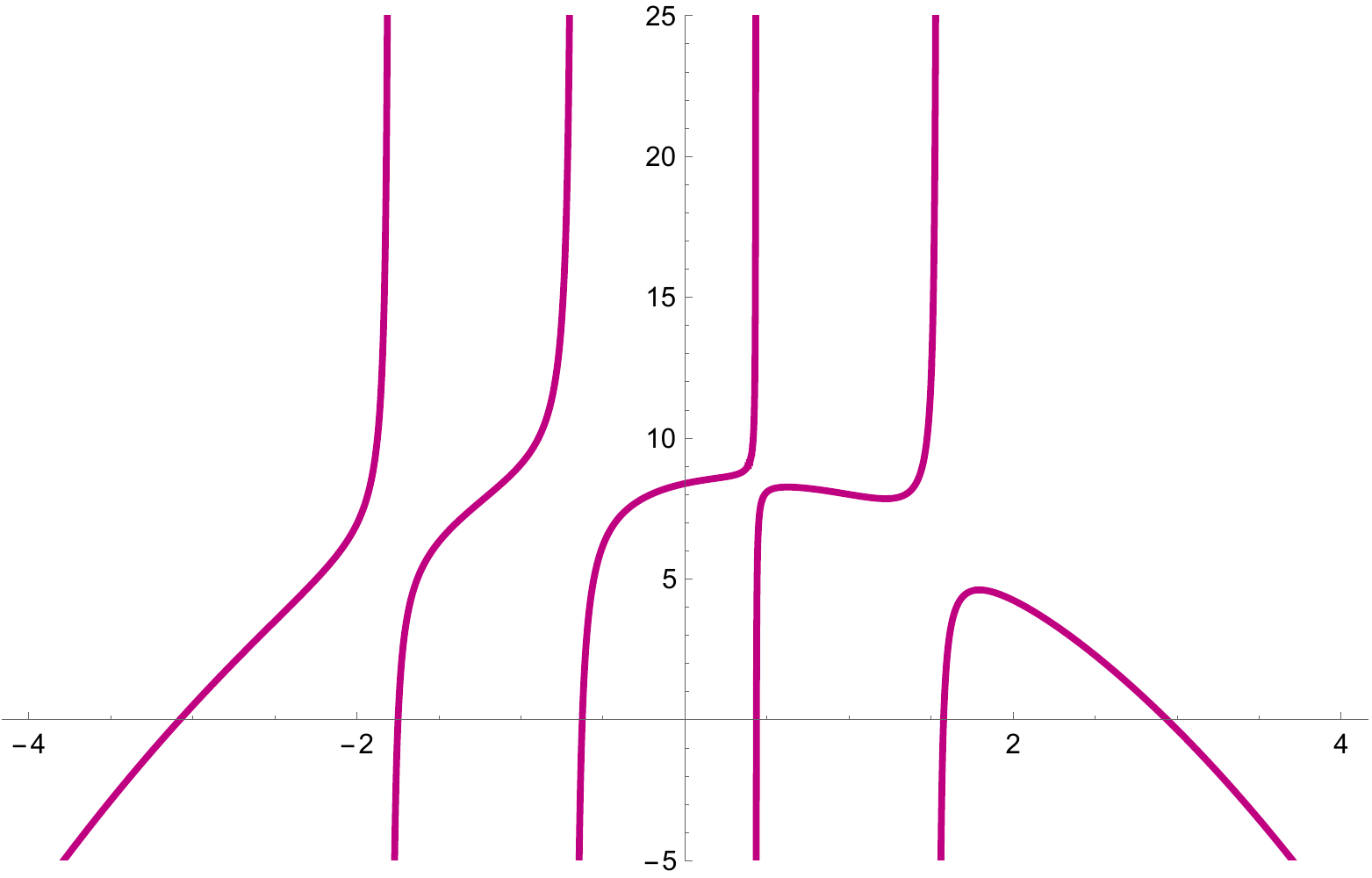}}
    \put(62,30){$V_Q^{(0,3,4)}$}
   \put(120, 0){\includegraphics[width=55mm]{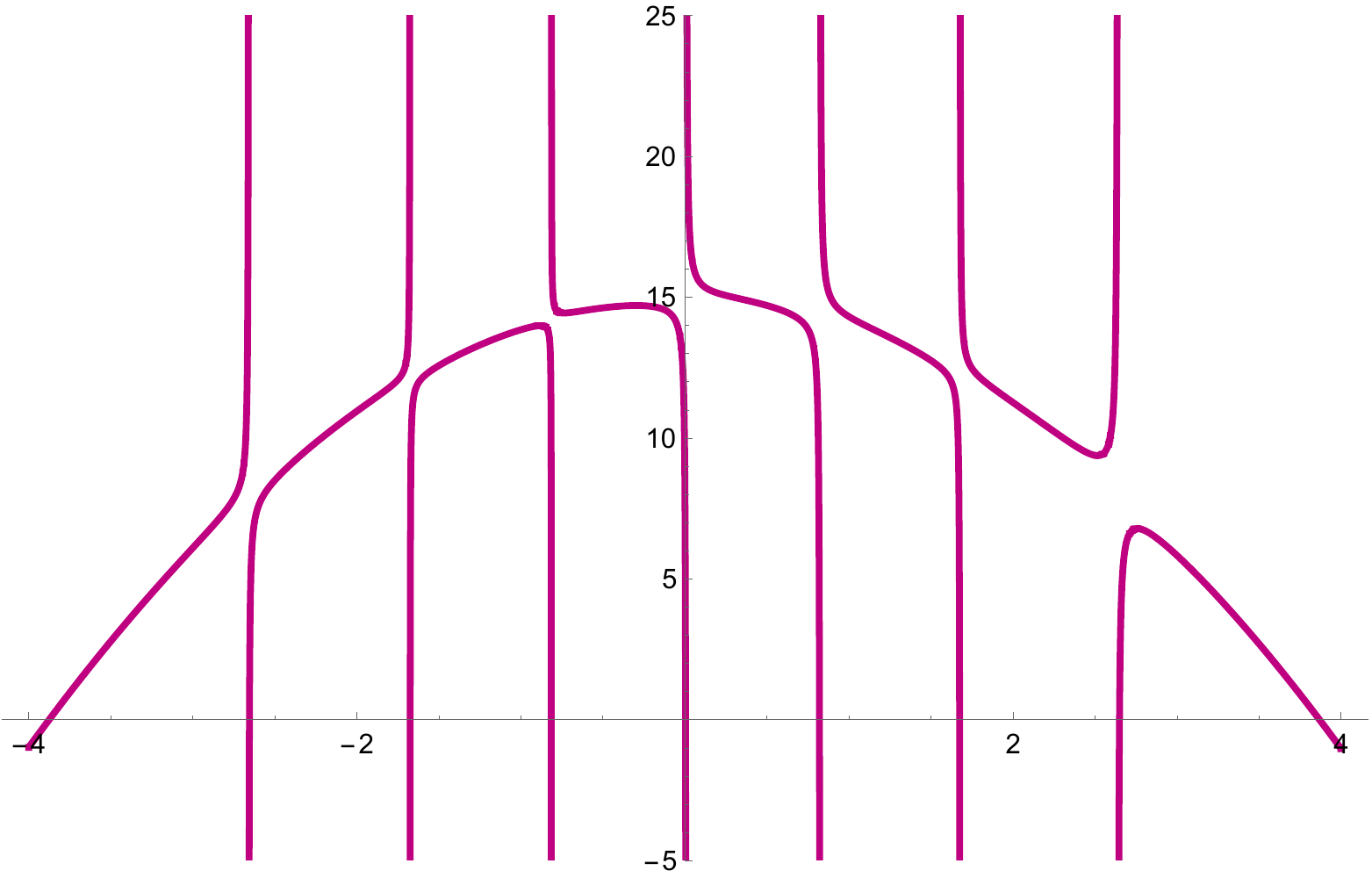}}
    \put(117,30){$V_Q^{(0,3,4,7)}$}
  \end{picture}
 \end{center}
 \caption{The ``quantum potential,'' $V_Q(\Psi)$ as defined in~\eqref{VQ}, plotted for a few linear combinations of lowest-energy states, illustrated here for
 $\Psi=\psi_0{+}\psi_3$, $\Psi=\psi_0{+}\psi_3{+}\psi_4$ and $\Psi=\psi_0{+}\psi_3{+}\psi_4{+}\psi_7$, in terms of harmonic eigenstates, $\psi_n$.}
 \label{f:VQ-Mix}
\end{figure}

In the harmonic approximation and for each harmonic eigenstate, we have in general that
 $\nabla^2\psi_n\propto\psi_n$,
producing the simple $V_Q^{(n)}=V_Q(\psi_n)$ computed from~\eqref{VQ}, and plotted in Figure~\ref{f:VQ-Hermite} for a few lowest-energy states.
In turn, the quantum potential computed from a general state (always representable as a linear combination of eigenstates), necessarily diverges at multiple poles; see Figure~\ref{f:VQ-Mix} for a few sample sketches.
In fact, this behavior stems from the definitions following standard quantum mechanics, with $\rho_\Psi\overset{\sss\rm def}=|\Psi|^2$:
\begin{equation}
  H \overset{\sss\rm def}= -\frac{\mockbar^2}{2m}\nabla^2+V(Q)
  \qquad\Leftrightarrow\qquad
  V_Q(\psi) \overset{\sss\rm def}=
   \frac1{\sqrt{\rho_\Psi}}\Big(-\frac{\mockbar^2}{2m}\nabla^2\sqrt{\rho_\Psi}\Big)
  =\frac1{\sqrt{\rho_\Psi}}\Big(\big[H-V(Q)\big]\sqrt{\rho_\Psi}\Big).
 \label{e:genVQ}
\end{equation}
Thus, when computed from any $H$-eigenfunction, $H\psi_n=E_n\psi_n$, it is immediate that
 $V_Q(\psi_n)=E_n{-}V(Q)$.
In fact, this holds generally --- provided the ``quantum potential'' is (re)defined following~\eqref{e:genVQ}:
\begin{alignat}9
  \widetilde{H} \overset{\sss\rm def}= K(P) +V(Q)
  \qquad&\Leftrightarrow&\qquad
  \widetilde{V}_Q(\Psi) &\overset{\sss\rm def}=
   \frac1{\sqrt{\rho_\Psi}}\Big(K(P)\sqrt{\rho_\Psi}\Big)
  =\frac1{\sqrt{\rho_\Psi}}
    \Big(\big[\widetilde{H}-V(Q)\big]\sqrt{\rho_\Psi}\Big),\\
  &~\text{so}&\qquad
  \widetilde{H}\widetilde\psi_n&=E_n\widetilde\psi_n \quad\Rightarrow\quad
  \widetilde{V}_Q(\widetilde\psi_n)=E_n-V(Q).
 \label{e:GenVQ}
\end{alignat}
We have also used that the $H$- and $\widetilde{H}$-eigenstates, $\psi_n$ and 
$\widetilde\psi_n$, are standing waves and have a constant phase (they have no probability current). 
For the case of the Lotka-Volterra Hamiltonian~\eqref{hlv}, the corresponding quantum potential should be defined as
$V_Q(\Psi)\overset{\sss\rm def}
=\frac1{\sqrt{\rho_\Psi}}\,d(e^P{-}P)\sqrt{\rho_\Psi}$, 
whereupon the result~\eqref{e:GenVQ} follows immediately.

This implies the following general statement:
In the ``landscape'' of possible states of the system, i.e., their probability distributions, $\rho_\Psi$, of a system with the Hamiltonian $H$, the (appropriately defined) quantum potential $V_Q(\Psi)$ merely changes the overall energy at eigenstates of $H$, but radically alters the potential (and so the dynamics) of the system away from those $H$-eigenstates.

The most important point is that, generically, we are led to a state-dependent dynamics and a new hierarchy of LV models, because the quantum potential, in general, changes with the mock quantum state. This is consistent with some global expectations in quantum biology \cite{davies}.
We emphasize that the robust predictions of mock quantum theory are the vacuum ``energy'' and the quantization of ``energy'' levels together with an emergent superposition of states and the Born rule for probabilities consistent with the superposition principle. These robust predictions should be found in the appropriate experimental realization of the LV model in its fully ergodic and far from equilibrium limit.
This discussion could be applied in other related situations \cite{latham}.

\subsection{Vacuum of the full MQ LV theory}
As promised, let us also examine the vacuum state of the full LV Hamiltonian
\be
H_{LV} = a \big(e^Q - Q\big) + d \big(e^P -P\big),
\ee
where $P=-i\mockbar\pa_Q$.
By Taylor's theorem, $e^{P}f(Q)=e^{-i\mockbar\pa_Q}f(Q)=f(Q-i\mockbar)$, which implies an analytic continuation of the {\it a~priori} real variable $Q$.
 We seek an eigenfunction of $H_{LV}$ in the exponential form for convenience:
\begin{equation}
 \begin{aligned}
  \hat{H}_{LV}\,e^{f(Q)}
  &=\big[a(e^Q-Q) +d(e^{\hat{P}}-\hat{P})\big]e^{f(Q)},\\
  &= \underline{a\,e^{Q+f(Q)}} 
    -a\,Qe^{f(Q)}
    +\underline{d\,e^{f(Q-i\mockbar)}}
    +i\mockbar d\,(\pa_Qf(Q))\,e^{f(Q)}
  =E\,e^{f(Q)}.
 \end{aligned}
 \label{e:HLV}
\end{equation}
 To insure the last equality for all $Q$, we compare terms that are functionally alike and conclude that the function $f(Q)$ must satisfy the conditions:
\begin{alignat}9
 \text{functional:}&\quad&
  f(Q-i\mockbar) &= Q+f(Q)+h, \label{e:fn}\\
 \text{differential:}&\quad&
  i\mockbar\,d\,(\pa_Qf(Q)) &= a\,Q+E, \label{e:df}
\end{alignat}
for some constant $h$, so that
\begin{alignat}9
  \hat{H}_{LV}\,e^{f(Q)}
  &= \underline{a\,e^{Q+f(Q)}}
    -a\,Qe^{f(Q)} 
    +\underline{d\,e^{Q+f(Q)+h}} 
    +i\mockbar d\,(\pa_Qf(Q))\,e^{f(Q)},\\
  &= \underline{(a+d\,e^h)e^{Q+f(Q)}}
    +E\,e^{f(Q)},
\end{alignat}
so that one must also require
\begin{equation}
 \text{consistency:}\quad
  a+d\,e^h=0. \label{e:cdh1}
\end{equation}

Now, both~\eqref{e:fn} and~\eqref{e:df} imply that $f(Q)$ must be a quadratic function, $f(Q)={-}\frac{(Q-Q_0)^2}{2\sigma^2}+i\varphi$.
Expanding~\eqref{e:fn} with this Ansatz and equating like terms in $Q$ implies that
\begin{equation}
  \sigma=\sqrt{i\mockbar},\quad
  Q_0=-\Big(h{+}\frac{i\mockbar}2\Big),\quad\text{and}\quad
  f(Q)=\frac{i}{2\mockbar}\Big[Q+\Big(h{+}\frac{i\mockbar}2\Big)\Big]^2 +i\varphi,
\end{equation}
where $h,\varphi$ remain free so far.
The differential condition~\eqref{e:df} then requires
\begin{alignat}9
 i\mockbar d\Big(\frac{i}{\mockbar}
                               \Big[Q+\Big(h{+}\frac{i\mockbar}2\Big)\Big]\Big)
 =-d\Big[Q+\Big(h{+}\frac{i\mockbar}2\Big)\Big]
 &\overset!=a\,Q+E,
\end{alignat}
where ``$\overset!=$'' denotes a required equality, so that
\begin{equation}
  d\overset!=-a, \quad\text{and}\quad
  E=-d\Big(h{+}\frac{i\mockbar}2\Big)
     =a\Big(h{+}\frac{i\mockbar}2\Big). \label{e:E0}
\end{equation}
Finally,~\eqref{e:cdh1} becomes $a(1-e^{h})=0$, which implies $h=h_n=2ni\pi$, and so $E=E_n=ia(\mockbar/2{+}2n\pi)$ for $n\in\mathbb{Z}$.

\medskip
To summarize, we have computed
\begin{equation}
 H_{LV}\,\psi_n(Q)  = E_n\,\psi_n(Q),\qquad
 E_n=ia\Big(\frac12\mockbar{+}2n\pi\Big),
\end{equation}
which required restricting $d=-a$ in $H_{LV}$ --- and so would impose an (unrealistic) condition on the LV system. Furthermore, the so-obtained eigenfunctions,
\begin{equation}
 \begin{aligned}
  \psi_n(Q)
 &=\exp\Big\{\,\frac{i}{2\mockbar}\Big[Q+i\Big(2n\pi{+}\frac12\mockbar\Big)\Big]^2
               +i\varphi\,\Big\},\\
 &=\exp\Big\{\,{-}\Big(\frac12{+}\frac{2n\pi}{\mockbar}\Big)Q \,\Big\}\;
   \exp\Big\{\, \frac{i}{2\mockbar}Q^2
               -i\Big[\Big(\sqrt{\frac{\mockbar}8}{+}\frac{2n\pi}{\sqrt{2\mockbar}}\Big)^2
               -\varphi\,\Big]\,\Big\},
 \qquad n\in\mathbb{Z}
 \end{aligned}
 \label{e:psin}
\end{equation}
are clearly not square-integrable: the real exponential factor diverges for $Q=\log(N_2/q_2)\to-\infty$, i.e., for $N_2\to0$. In turn, the standard (quadratic approximation) quantum potentials computed from~\eqref{e:psin},
\begin{equation}
  V_Q[\psi_n(Q)]=-\frac12 d\mockbar^2\frac{\pa_Q\!^2|\psi_n(Q)|}{|\psi_n(Q)|}
  =-\frac18 d(4n\pi+\mockbar)^2,
\end{equation}
are simple $n$-dependent constants. The ``LV-exact'' quantum potentials computed from~\eqref{e:psin}
\begin{equation}
  V_Q^{LV}[\psi_n(Q)]=d\frac{[e^{\hat{P}}{-}\hat{P}]|\psi_n(Q)|}{|\psi_n(Q)|}
  =\frac12 d\big( 2 e^{\frac{i\mockbar}2}-i (4n\pi+\mockbar) \big)
  =-a e^{\frac{i\mockbar}2} +iaE_n,
\end{equation}
are also $n$-dependent, but {\em\/complex\/} constants, typically encoding dissipation.

 The functions in the sequence~\eqref{e:psin} may perhaps be thought of as the initial ($\lambda=0$) perturbative solutions (presumed to be analytic functions of $\lambda$) of the non-linear extension
\begin{subequations}
 \label{e:QMLV}
\begin{gather}
  \bigg[ H_{LV}^Q \mathrel{\overset{\sss\text{def}}=} 
         a\big(e^Q-Q\big) +d\big(e^{\hat{P}}{-}\hat{P}\big) 
         +\lambda\underbrace{d\frac{\big((e^{\hat{P}}{-}\hat{P})|\Psi_n(Q;\lambda)|\big)}
                                  {|\Psi_n(Q;\lambda)|}}_{V_Q[\Psi(Q;\lambda)]} \bigg]
  \Psi_n(Q;\lambda) = E_n(\lambda)\,\Psi_n(Q;\lambda), \\
  \Psi_n(Q;\lambda)=\sum_k\lambda^k\Psi_n^{\sss(k)}(Q),\qquad
  E_n(\lambda)=\sum_k\lambda^kE^{\sss(k)}_n.
\end{gather}
\end{subequations}

The excited states of the full mock LV system might be hard to write explicitly, but that does not change the fact that the mock LV dynamics is generically state-dependent. This is one of the most important points of the above discussion. This state dependent nature of the mock quantum LV dynamics should be contrasted with the state dependent dynamics of classical non-linear systems \cite{latham}.
The full mock quantum LV equations, written in terms of the original $N$ variables, change with the mock quantum state, which is the central new feature of our discussion
(and which fits into the discussion of \cite{davies} and references therein).

\section{Stochastic Dynamics and Mock Quantum Statistical Field Theory}
\label{s:7}
As advertised in Section~\ref{s:2}, now we discuss the issue of imperfect cancellation between the quantum potential and the state-dependent environment, by considering more realistic stochastic dynamics.
We will show that stochastic dynamics can be easily incorporated in our formalism and that in general, stochastic mock quantum dynamics leads to a state-dependent statistical field theory.

It is well known that the stochastic LV dynamics (described by the Langevin equations) can be converted into
statistical field theory; see for example \cite{michel}.
In this section we examine the mock-quantum version of statistical field theory.
Let us denote $N_1$ and $N_2$ in the LV system as $\phi_i$, where $\phi_i$ (functions of time $t$) satisfy the
Langevin equation  \cite{michel}
\be
\lambda^{-1} \partial_{t} \phi_i = F_i[\phi_i] + \xi,
\ee
where $\xi$ is the Langevin noise with the following noise correlator
\be
\langle \xi(t)\, \xi(t') \rangle = \lambda^{-1} N [\phi_i]\, \delta(t - t').
\ee
(In the simplest case $N=1$, and $F$ is defined from the deterministic system of LV equations.
Also, $F_i [\phi_i] \define -\frac{\delta H(\phi_i)}{\delta \phi_i}$, where $H$ is in general the Landau-Ginzburg (Hamiltonian) functional.)
The trick is now to write the following partition function
\be
Z = \int \rD\xi~ P(\xi) \int \rD\phi_i~ 
\delta( \lambda^{-1} \partial_{t} \phi_i - F_i[\phi_i] - \xi ),
\ee
where the Langevin noise is assumed to have a Gaussian distribution
\be
P(\xi) \sim \exp\Big({-}\!\int \rd t~ \frac{\xi^2}{2 k N} \Big).
\ee
Then by using the identity
\be
\delta(x) = \frac{1}{2 \pi} \int \rD \tilde{\phi_j}~ \exp(i \tilde{\phi_j} x ),
\ee
we obtain the following partition function (after the appropriate Wick rotation in the complex plane)
\be
Z \sim  \int \rD\xi~ P(\xi)\, e^{-\int dt \phi_j \xi } 
\int \rD\phi_j\, \rD \tilde{\phi_j}
e^{- \int \rd t \tilde{\phi_j}(\lambda^{-1} \partial_{t} \phi_j - F_j[\phi_j]) }.
\ee
Finally, we perform the Gaussian integration over $\xi$ to obtain
\be
Z \sim \int \rD\phi_j\,\rD\tilde{\phi_j}~ e^{- J(\phi_j, \tilde{\phi_j} ) },
\ee
where the effective action $J$ of this statistical field theory reads as
\be
 J(\phi_j, \tilde{\phi_j} ) 
 = \int \rd t~\tilde{\phi_j}
    \Big( \lambda^{-1} \partial_{t} \phi_j -F_j[\phi_j] 
          -\frac{k}{2} N[\phi_j] \tilde{\phi_j} \Big) .
\ee

Now, the difference between the canonical statistical field theory, and the mock quantum version of statistical field theory, is in the form of $F$: 
It follows from our previous section that $F$ changes (by an addition of the term generated from the state
dependent quantum potential) between the classical and mock quantum form of the LV equations.
In particular, in the mock quantum (MQ) case
\be
F_i [\phi_i] \define -\frac{\delta H(\phi_i)}{\delta \phi_i} ~\to~
F^{MQ}_i [\phi_i] \define -\frac{\delta H(\phi_i) +\delta V_Q (\phi_i)}{\delta \phi_i}.
\ee
Thus the mock quantum statistical field theory is defined by the following mock quantum partition function
\be
Z^{MQ} \sim \int \rD\phi_j\, \rD \tilde{\phi_j}~
e^{- J^{MQ}(\phi_j, \tilde{\phi_j} ) },
\ee
where the effective action $J^{MQ}$ of this mock quantum (and {\em state-dependent\/}!) statistical field theory reads as
\be
 J^{MQ}(\phi_j, \tilde{\phi_j} ) 
 = \int \rd t~ \tilde{\phi_j}
    \Big( \lambda^{-1} \partial_{t} \phi_j -F^{MQ}_j[\phi_j] 
          -\frac{k}{2} N[\phi_j] \tilde{\phi_j} \Big).
\ee

Such state-dependent statistical field theory is one way of including the external/environment noise in our the context of mock-quantum theory.
Thus the cancellation/balancing between the state-dependent quantum potential $V_Q(\psi, \psi^{\dagger})$ and 
the state-dependent environment term 
$\eta(\psi, \psi^{\dagger})$
does not have to be perfect, and their non-vanishing remnant can be included using the above formalism.

\section{Why $V_Q$?: complexity, variety and mock quantum theory}
\label{s:8}
In the preceding discussion one of the most important ingredients was the mock quantum potential $V_Q$.
The inclusion of $V_Q$ gives crucial new results even in the quadratic (harmonic) approximation of the mock quantum LV model:
a) first, one gets a nonzero vacuum energy, and 
b) one gets discrete ``energy'' states.
Also, in general, one has interference between different states.
Even in the quadratic limit these two features (as well as interference) should be observed in real-world applications (for example in the context of the Wilson-Cowan model, which is a computational neuroscience model which describes dynamics of populations of interacting neurons \cite{latham}), if indeed mock quantum theory is relevant for real complex adaptive systems.
More fundamentally, the form of $V_Q$ is state dependent and thus in mock quantum theory we have a realization of ``state-dependent dynamics'' (usually realized in classical, stochastic, network dynamics \cite{latham})
but with crucial quantum-like features.

In Section~\ref{s:4} we have given a generic argument for why the non-equilibrium dynamics of a classical, complex adaptive system might lead to a mock quantum description which crucially involves the mock quantum potential.
However, the astute reader might rightfully ask: What is the deeper physical meaning of $V_Q$ and why would such
a non-classical potential appear in the
classical dynamics of complex adaptive systems?
(In particular, the LV model might involve ``rabbits and foxes'' or ``sharks and tuna'' \cite{lv}, and so it
seems almost fantastic that a ``mock quantum potential'' would have anything to do with such macroscopic systems.)
In this section we want to shed light on this very important question.

We will elucidate this question by using the recent work of Smolin \cite{smolin}
(originating with Barbour and Smolin \cite{barbour})
on the connection between complexity and variety and the quantum (Bohmian) potential.
In some sense, the quantum potential is (according to Smolin) a measure 
of the variety of
a collection of similar subsystems of a given system.
(In his work \cite{smolin} Smolin aims at interpreting the fundamental quantum potential and he attempts to derive fundamental
quantum theory with ultimate applications to cosmology. This is not our present aim. However, we will use Smolin's work
to elucidate the emergence of mock quantum theory in classical complex adaptive systems in which the
adaptation of a system to the environment is crucial.)
The central reason of why invoke this work is because of its emphasis on the principle of ``maximal variety'', which is very natural in the context of complex adaptive systems, and its relation to the quantum potential.

Our reading of Smolin's suggestion is as follows:
the mock quantum potential is a measure of the variety of responses a complex adaptive system has
in a given environment. The principle of maximal variety leads to the expression for the mock quantum potential
that we used in previous sections.

Thus, in the context of ``predator-pray'' LV systems in a given environment, the
principle of maximal variety of adaptive response these systems have in their environments, could
lead, in principle, to the mock quantum potential, and mock quantum theory.
In the case of these macroscopic systems the required time scale might be unrealistic for
observing the emerging mock quantum behavior, and the balancing between the mock quantum potential and the environment might be impossible.
However, the network dynamics of neurons (modeled by the Wilson-Cowan (WC) model \cite{latham} related to the LV model)
or the effective LV dynamics in cellular environments may well lead to more realistic emergent mock quantum systems.

Here we summarize the derivation of the quantum potential from the principle of maximal variety 
as outlined in Smolin's papers \cite{smolin} (see also \cite{barbour}).
First Smolin defines the concept of variety following Barbour: ``The variety of a system of relations, $\cal{V}$, is a measure of how easy it is to distinguish the
neighborhood of every element of a system from that of every other.''
(Note that what is assumed here is that there are ``many interacting classical systems,'' which is appropriate in the case of mock
quantum theory.)

The elements of the system, might be particles or other classical entities, and relations (or relational observables $X_{ij}$)
between these elements:
for example, particles and their relative distances. Smolin also defines the view of the $i$-th element, which is what element $i$ may ``know''
via relational observables $X_{ij}$ about the rest of the system. The view is denoted by $V_i(X_{ij})$.
Also, the distinctiveness of two elements, $i$ and $j$ is a measure of the differences
between the views of $i$ and $j$: 
\be
I_{ij} \define |V_i-V_j|^2,
\ee
and it acts as a natural metric on the set of subsystems (two systems are close if they have similar views).
Then the variety is defined to measure the distinguishability of all the elements (the total number being $N$) from each
other
\be
{\cal{V}} = \frac{1}{N(N-1)} \sum_{i \neq j} I_{ij} = \frac{1}{N(N-1)} \sum_{i \neq j} |V_i-V_j|^2.
\ee

Smolin \cite{smolin} then looks at a continuous form of this expression,
by introducing the probability density $\rho$ and a fundamental cut-off determined in terms of the inverse
of the probability density.
Essentially Smolin \cite{smolin} replaces the above discrete formula with a continuous form via the following dictionary:
\be
\frac{1}{N} \sum_k \phi(x_k) \to \int \rd^d z~ \rho(z)\, \phi(z),
\ee
as well as
\be
\frac{1}{N} \sum_k \phi(x_{k+i}, x_k) \to \int \rd^d x~ \rho(z+x)\,\phi(z+x, z).
\ee
Then the continuous formula for the variety ${\cal{V}}$  involves the product of probability densities \cite{smolin}
\be
{\cal{V}} \to \int \rd^d z~ \rho(z) \int \rd^d x~ \int \rd^d y~
               \rho(z+x)\,\rho(z+y),
\ee
which upon the Taylor expansion of $\rho(z+x)$ and $\rho(z+y)$
leads to the following result for the variety \cite{smolin}
\be
{\cal{V}} \to \int \rd^d z~ \rho\,\Big(\frac{1}{\rho}\,\partial_a \rho\Big)^2 +...
\ee
which ultimately gives the quantum potential of the de-Broglie-Bohm type. Smolin \cite{smolin} relates the potential
energy to this expression and claims that this quantum potential is the leading term that comes from maximal variety.
He then goes on to derive the Schr\"{o}dinger equation from this principle of maximal variety \cite{smolin}.

This approach is eminently suitable for applications to complex adaptive systems, which are macroscopic
classical systems far from equilibrium, and one can thereby envision the application of the principle
of maximal variety as defined by Smolin and Barbour.
The above derivation of the quantum potential then becomes applicable to complex adaptive systems
and thus makes our claim regarding the emergent quantum (mock quantum) description more robust.
One only needs to substitute the canonical Planck constant (set to one in the above discussion) with a non-universal and system-dependent mock Planck constant $\mockbar$.
Thus, the mock quantum formulation has the required robustness, at
least on the level of theory. 
Note that the previous discussion is also consistent with an empirical view that data describing complex systems do not always conform with the laws of classical probability, but rather with quantum-like probability \cite{plotnitsky,khrenikov},
which is consistent with a more general picture of quantum mechanics as quantum measure theory
\cite{sorkin}.
Nevertheless, it is crucial to find some experimental evidence for mock quantum dynamics. This brings us back to the hydrodynamic example mentioned as one of the original motivations for our work \cite{droplet, fluidpilot}.

\section{Emergent Hydrodynamics: the (Mock) Quantum-Classical Transition}
\label{s:9}
Here we use the above discussion of ``why $V_Q$?" in order to point out a
very useful hydrodynamic formulation of Mock Quantum Theory, following the original insight of E.~Madelung~\cite{madelung1927}.
(As a reference for this discussion see \cite{holland}.)
In this context (which connects our work to the ongoing discussions found in \cite{droplet, fluidpilot}) we also discuss some universal features of the canonical quantum-to-classical and also mock-quantum-to-classical transitions.

In Appendix~\ref{a:C} we summarize the relatively standard derivation, where starting from the de-Broglie-Bohm formulation of classical Hamilton-Jacobi equations (which includes $V_Q$), we get 
the ``(mock) quantum Euler'' equation
\be
\frac{\partial v_i}{\partial t} + (\vec{v} \cdot \nabla) v_i = -\frac{1}{m} \partial_i V - \frac{1}{m \rho} \partial_j \sigma_{ij},
\label{sig.ij}
\ee
where the ``(mock) quantum'' stress-tensor $\sigma_{ij}$ reads
\be
\sigma_{ij} \define 
-\frac{\mockbar^2 \rho}{4m}\,\partial_i\partial_j \log{\rho}.
\label{s.ij}
\ee

We stress the crucial role of the (mock) quantum stress-tensor $\sigma_{ij}$
 or, equivalently, the (mock) quantum potential, $V_Q$, shown above.
The classical limit corresponds to turning off the (mock) quantum stress-tensor term, which leads to the Euler equation-like limit of ``(mock) quantum hydrodynamics.'' Alternatively, the mock quantum dynamics is induced by turning on the (mock) quantum stress-tensor, by placing the classical system in a very particular environment that is effectively modeled by the (mock) quantum stress-tensor or, equivalently, (mock) quantum potential. In what follows, we argue that in the limit of the ``turbulent'' phase of the above mock quantum hydrodynamics one may identify the observable effects of the quantum-to-classical and mock-quantum-to-classical transitions.

The following observations are in order: 
The discussion in the previous Section relates the (mock) quantum potential $V_Q$ to the principle of maximal variety
(and thus maximal variety of adaptations of the system under consideration to its environment),
which then leads to the above ``mock quantum'' stress tensor, 
and thus to the above hydrodynamic formulation.
Note also: $\rho$ measures probability, and the logarithm of probability is entropy by the Boltzmann formula,
whereupon the (mock) quantum stress tensor is related to the second derivative of entropy, which for maximally symmetric
(Fisher) metric in the space of probabilities \cite{wootters, chia}, is given by the robust features of the Gaussian distribution that is related to the Fisher metric.
This again confirms the robustness of the above hydrodynamic formulation and also robustness of Mock Quantum theory,
which emerges from the maximal variety of adaptation of a complex system to its environment, by the discussion in this and the previous Section.

This hydrodynamic formulation should be related to the ongoing discussion on the validity of the actual microfluidics experiments that might be illustrative of an 
emergent ``quantum'' behavior
in macroscopic systems \cite{droplet, fluidpilot}. The most natural interpretation of
these experiments (assuming their ultimate validity) is in terms of emergent or
mock quantum theory.
The mathematical models used to describe these
phenomena could be used in different contexts
(such as the area of complex adaptive systems),
but the models used in \cite{fluidpilot}
seem to be fine-tuned and perhaps they are not as
useful in the context of biology.
However, our claim that the same (Mock Quantum) outcomes that are claimed to be observed in these experiments
\cite{fluidpilot}
should be realized 
in complex adaptive biological systems, but more
generically, given the above argument regarding the emergence of the (mock) quantum potential 
from the principle of maximal variety, which
is particularly well suited for such systems.

Here we make an important observation regarding the universality of the quantum-to-classical and mock-quantum-classical transitions which also relies on the central role of the mock quantum stress-tensor or the mock quantum potential.
The following discussion also illustrates the crucial distinction between the classical and the mock-quantum dynamics in the hydrodynamic formulation.

It has been suggested by Leggett \cite{tony} that in macroscopic
quantum systems one might have to deal with many entangled
objects (not, say, the pairwise entanglement found in superconductors, for example), and that for a critical 
number of entangled objects, one might have a situation
analogous to the turbulent phase of fluid dynamics, where above
the critical Reynolds number (the ratio of the
non-linear $(\vec{v} \cdot \nabla) v_i$ and the viscous term in the Navier-Stokes equations~\eqref{(v.d)v} in Appendix~\ref{a:B})
one encounters a very
different (turbulent) physics behavior as opposed to the physics
found below the critical Reynolds number.

Obviously in the above discussion of ``quantum hydrodynamics'' one could consider 
a (mock) ``quantum Reynolds number,'' the ratio of the non-linear 
term $(\vec{v} \cdot \nabla) v_i$ and the relevant term that depends on the (mock) quantum stress tensor
$\sigma_{ij}$ in~\eqref{sig.ij}, Appendix~\ref{a:B},
so that above a critical number (in the phase where the non-linear term $(\vec{v} \cdot \nabla) v_i$ 
dominates over the $\mockbar$- (or $\hbar$-)dependent quantum stress tensor term) one gets ``(mock) quantum-to-classical turbulence'' characterized
by a non-trivial (for example, Kolmogorov-like \cite{Gray:2013rv}) scaling of correlations functions of ``velocities''
$v_i$. The same statement would be true for mock quantum theory
in which the Planck constant is replaced by the effective (mock) Planck constant.

In particular, we could apply the Kolmogorov insight \cite{Gray:2013rv}
for
the flow of ``energy'' $v^2$ (where $v \sim \nabla S/m$,
that is, the effective velocity is the gradient of the phase of the wavefunction) 
\be
v^2/t \sim \epsilon \to v \sim (\epsilon\,l)^{1/3},
\ee
where we have used the scaling $v \sim l/t$, with $t$ denoting
the time evolution variable and $l$ the characteristic spatial scale.
In that ``turbulent" regime the effective ``velocity,'' and thus $S$ (that is, (mock) quantum phases) become 
random fields with the following two-point correlator, as
implied by the above scaling
\be
\langle v(l)\, v(0) \rangle \sim 
\big\langle (\nabla S/m)(l)\, (\nabla S/m)(0) \big\rangle 
\sim (\epsilon l)^{2/3} \to
\langle S(l)\, S(0) \rangle \sim (\tilde{\epsilon}\,l)^{8/3}.
\ee
This would be a universal feature of the quantum-to-classical and also of the mock-quantum-classical transition that could, in principle, be observed experimentally. Note that in this context the quantum limit corresponds to the ``laminar flow,''
whereas the classical dynamics corresponds to the ``turbulent flow'' of the gradient of the phase of the wavefunction.

\section{Conclusion} 
\label{s:10}
In this paper we have elaborated on the proposal of \cite{Minic:2014zsa}
that was motivated by certain fluid dynamics experiments reviewed in \cite{droplet, fluidpilot}.
It is still an open question whether such experiments (if, indeed, reproducible) can 
be repeated in the context of living systems.
Nevertheless,
we have herein explored the concept of emergent quantum-like theory in complex adaptive systems.
We have also presented a generic reason why the non-equilibrium dynamics of a classical complex adaptive system would lead to a mock quantum description, something that was not present in the original proposal.
In particular, we have 
examined the concrete example of emergent (or mock) quantum theory in the Lotka-Volterra system.
In this context, we have emphasized the state-dependent nature of the mock quantum dynamics and we have introduced the new concept of
mock-quantum, state-dependent, statistical field theory.
We have also discussed some universal features of the quantum-classical as well as the mock-quantum-classical transition
found in the ``turbulent'' phase of the hydrodynamic formulation of our proposal.
Mock quantum theory might justify a purely empirical view that data describing complex systems do not always conform with the laws of classical probability, but rather with quantum-like probability \cite{plotnitsky,khrenikov}. 

We note
that
oscillatory phenomena (of the kind used in \cite{droplet, fluidpilot}) are ubiquitous in biology, both at the
tissue and at cellular levels \cite{cellosc}, and are essential for the functioning of biological systems. It is then
tempting to conjecture that by investigating such oscillations, a potential evidence for the necessity and emergence
of mock-quantum stability can be demonstrated. While real biological systems might be too complicated to analyze, a
simplified synthetic regulatory cell or an auto-catalytic chemical process, in which the nature of these
oscillations can be controlled, could be a platform for investigating the relevance of our proposal \cite{synthetic}.
What we propose as a test of our theoretical discussion should essentially be seen as a direct analogue of the recent quantum-like fluid dynamics experiments \cite{droplet, fluidpilot}, however, to be conducted in the context of synthetic biology.

The formal quantization of the Lotka-Volterra system (using the usual Planck constant $\hbar$)
has been discussed previously in the literature \cite{enz}. However, that discussion is unrelated to the central point of this work regarding the emergence of an effective quantum theory in complex adaptive
systems.  We could also compare our proposal to the approach to quantization known as stochastic quantization
\cite{parisi}, and state that such developments (even though they are made in a completely different context of quantum foundations) are illustrative of our claim that mock quantum dynamics is achievable in macroscopic complex adaptive systems.

The ``mock quantum theory'' proposal can be
intuitively understood as a new type of non-classical adaptive stability in biological systems, but should be clearly distinguished from
the arguments that canonical quantum physics is relevant in biological systems \cite{quantumbio}. 
Our effective, mock-quantum theory comes with a new fundamental deformation parameter (e.g., the size of the action of the
Lotka-Volterra models) that is emergent and so distinguishable from the canonical fundamental quantum theory, which
usually suffers (due to decoherence) in competition with realistic thermal and noisy biological environments. 
The original article on mock quantum theory \cite{Minic:2014zsa} only
started the study of these emergent, mock-quantum phenomena \cite{supermq}, and we have aimed herein to explain in more detail the robustness of this proposal
as well as its various implications.

A central point of our present work is the state-dependence of the dynamics and a new hierarchy of models (such as the LV model), given the fact that the quantum potential changes with general mock-quantum states, and which is consistent with some general arguments in ``quantum biology''
\cite{davies}.
One intuition regarding the possible relevance of the mock-quantum proposal in complex adaptive systems is the competition
between stability and adaptability. 
Even though it is fair to say that classical stochastic complex systems are not completely understood,
one insight regarding them is that they
are neither too chaotic nor too ordered, but that they exist at some kind of a ``criticality'' \cite{bialek}.
Quantum-like fluctuations generically suppress chaos, but they increase
computability (intuitively, given the path integral formulation of 
quantum dynamics \cite{feynman}, quantum computers are exponentially enhanced
classical computers) and thus one would expect, at least very
naively, that in emergent quantum (or mock quantum) complex adaptive systems, chaos is suppressed. Their ``quantum-like nature'' might therefore enhance stability, and 
both from the information-theoretic point of view and because
of enhanced computability, such systems might be more adaptable.
Thus quantum-like effects might be utilized by evolution for the reasons of enhanced stability and adaptability.
Still, we should remember that standard quantum theory will
always face problems of decoherence and difficulty of scaling-up in generic macroscopic systems and that mock quantum theory,
while evading decoherence, seems to face the issue of 
canceling/balancing of the system under consideration to its complex adaptive (and, in general, stochastic) environment.
We hope to explore some of these fascinating issues in our future work.

{\bf Acknowledgments:}
We thank V.~Balasubramanian, D.~Bolmatov, N.~Goldenfeld, J.~Katsaras, M.~Pleimling, L.~Smolin, T.~Takeuchi, U.~Tauber, C.~Tze and V.~Vedral for comments.
We also acknowledge the insightful, constructive and detailed comments of the anonymous referees.
 TH is grateful to the Department of Mathematics, University of Maryland, College Park MD, and the Physics Department of the Faculty of Natural Sciences of the University of Novi Sad, Serbia, for the recurring hospitality and resources.
 The work of DM is supported in part by Department of Energy (under DOE grant number DE-SC0020262) and the Julian Schwinger Foundation, and the work of SP by the Division of the Intramural Research Programs (DIRP) of the National Institute of Mental Health (NIMH), USA, (project ZIAMH002797).

\appendix

\section{Hydrodynamic quantum-like behavior}
\label{a:Hydro}

\begin{figure}
    \centering
    \includegraphics[scale=0.67, viewport= 30 50 550 500, clip]{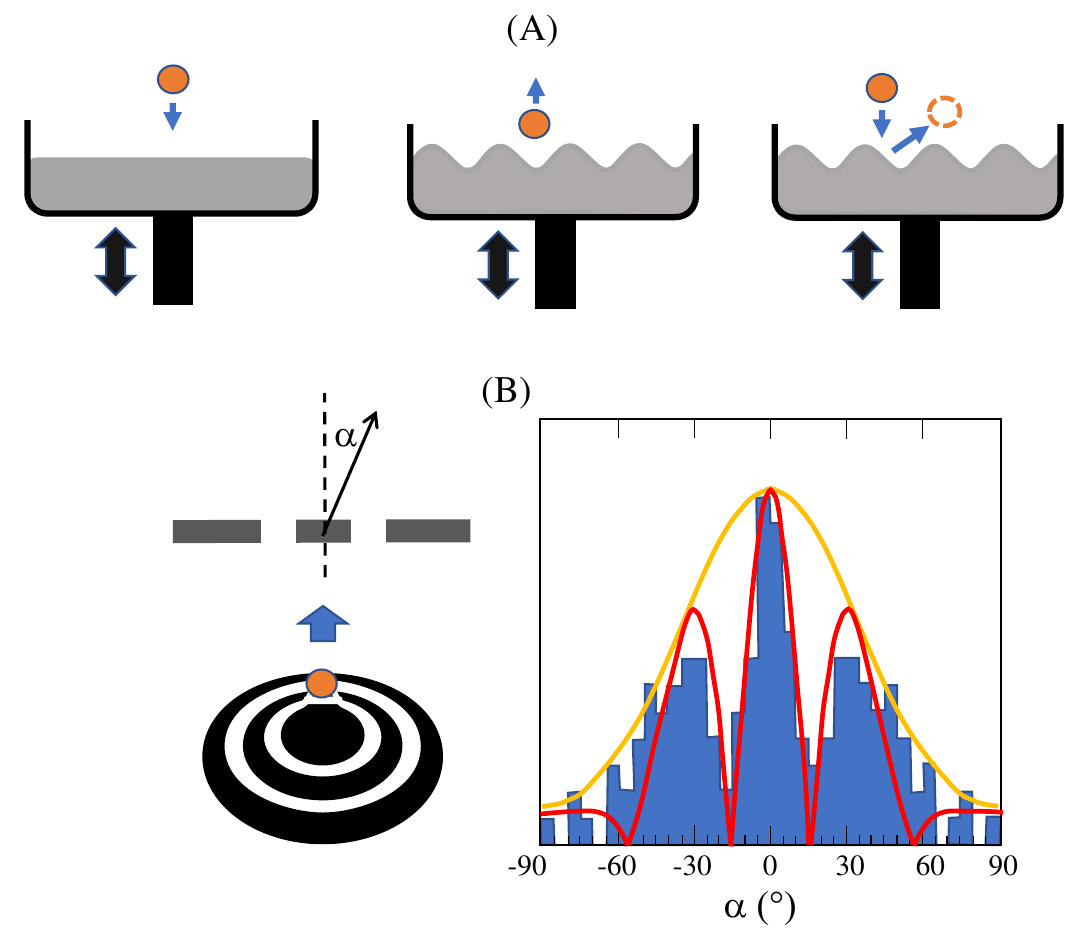}
    \caption{Quantum-like behavior of a millimeter size droplet. (A) Creation of a 'Walker': (1st, left panel) Bath with silicon oil is subject to vertical oscillations, and a drop falls on the surface.  (2nd, central panel) A drop has bounced of the fluid surface and created a transient wave.  (3rd, right panel) Walker: the drop continues bouncing but also moves horizontally because of the wave, which is a combination of the Faraday wave (due to bath vibrations) and a transient wave created by the impact of the droplet. Experiments are performed in near critical regime for Faraday waves when the surface is flat in absence of the drop. \\ 
    (B) Double-slit experiment: (left panel) A walker is moving in the direction of a double slit. (Right panel) The histogram of the deflection angle ($\alpha$) of a walker, which is consistent with the double-slit QM experiment  --- reconstructed from \cite{fluidpilot}. A wavelike histogram is emerging, with three peaks (red lines), approximately in agreement with a double-slit interference pattern for a monochromatic wave of the wavelength $\lambda_F$ of the Faraday wave.}
    \label{fig:Hydro}
\end{figure}

The "walking droplet" system is an interesting hydrodynamic system which can serve as a model for demonstrating quantum-like statistics \cite{droplet,droplet2, fluidpilot}. In the hydrodynamic experiments described in \cite{droplet,fluidpilot}, the droplet-wave system initially consists of two parts: (1) the Faraday wave in a fluid in a bath created by the vertical vibrations of the bath, (2) a millimeter size droplet over the surface of the fluid, Fig.~\ref{fig:Hydro}(A). The fluid surface is initially flat because the Faraday wave is in subthreshold but near critical regime, Fig.~\ref{fig:Hydro}(A), the first panel. Then the droplet falls on the fluid surface and bounces, and the impact creates a transient wave in the fluid, Fig.~\ref{fig:Hydro}(A), the second panel. Then the droplet falls on the fluid again, but now there is a surface wave which causes the droplet to bounce sideways, causing additional horizontal motion of the bouncing droplet, Fig.~\ref{fig:Hydro}(A), the third panel. This process can create a self-propelling droplet which can simultaneously bounce and move over the surface --- the so-called “walker”. For certain parameter regimes, walkers and quantum particles display similar {\bf long term statistical behavior}, with the fluid properties surface tension and density ($\sigma$, $\rho$) standing in for ($\hbar$, $m$). Initial experiments also claimed that this hydrodynamic particle/pilot-wave system can demonstrate certain features of quantum world, such as double-slit interference, tunneling, quantized orbits etc, see Fig.~\ref{fig:Hydro}(B). However, this observation has been disputed in a series of experiments by Bohr et al.~\cite{droplet_critique} and Pucci et al.~\cite{droplet_critique2}. A comprehensive analysis of this delicate hydrodynamic pilot-wave/particle systems is provided by Bush and Oza~\cite{Bush-Oza}. They describe many intricacies of the experiment including the features critical for the emergence of quantum-like behavior, such as association of the droplet with a self-propelling wave, the resonance between the particle and the wave, and some sort of temporal non-locality.

\section{Approaching Mock Quantum Theory}
\label{a:B}
In this appendix we present different (though equivalent) approaches to emergent (mock) quantum theory, which should be useful to keep in mind as we proceed with our analysis in the rest of the paper. 
We concentrate on the deterministic systems in this section, in order to make the general argument, and we postpone
the discussion of stochastic systems to Section~\ref{s:7}.
Perhaps the easiest way to think about mock quantum theory is by the following parallel with the classical theory that runs as follows
(we start with a textbook presentation \cite{feynman} in order to establish some basic notation).

Deterministic classical theory is well known to be defined via the variational principle (for simplicity, we showcase one $q$ variable, but a multi-variable description is analogous --- with but systematically replacing $q\to q_i$ and employing the Einstein summation convention), requiring
$\delta S= 0$,
where $S\define \int p\,dq - H(q, p)\,dt \define \int L (q, \dot{q})\,dt$ is Hamilton's action functional,
$H$ is the Hamiltonian, 
$L$ the Lagrangian, 
and $\dot{q} \define \frac{d q}{ d t}$.
Note that $S$ is a function of $q$: $S(q)$.
The vanishing variation of the action $\delta S=0$ implies the Euler-Lagrange equations:
\be
\delta S=0 ~\to~ \frac{\rd}{\rd t} \Big( \frac{ \partial L}{\partial \dot{q}}\Big)
 = \frac{\partial L}{\partial q}.
\ee
(This Lagrange picture becomes the path integral picture in the context of mock quantum theory.)
Note that the Noether theorem (generally relating
symmetries and conserved quantities) associated with the
independence of physics with respect to the time translation $ t \to t + \epsilon$ implies the Legendre-transform like relation
between the Lagrangian and the Hamiltonian. For if the Lagrangian does not implicitly depend on time, $\frac{\partial L}{ \partial t} = 0$, then
\be
\frac{\rd L}{\rd t} = \frac{\partial L}{\partial q} \frac{\rd q}{\rd t} + \frac{\partial L}{\partial \dot{q}} \frac{\rd\dot{q}}{\rd t} 
= \frac{\rd}{\rd t} \Big( \frac{ \partial L}{\partial \dot{q}}\Big) \dot{q} + \frac{\partial L}{\partial \dot{q}} \frac{\rd \dot{q}}{\rd t} 
= \frac{\rd}{\rd t} \Big(\frac{ \partial L}{\partial \dot{q}} \dot{q} \Big),
\ee
where we have used the Lagrange equations of motion, 
$\frac{\rd}{\rd t} ( \frac{ \partial L}{\partial \dot{q}}) = \frac{\partial L}{\partial q} $.
By defining the momentum,
$p \define \frac{ \partial L}{\partial \dot{q}}$, 
we get the conserved Hamiltonian, $\frac{\rd H}{\rd t} =0$:
energy is conserved because of the symmetry of time translation --- Noether's theorem.
From the variation of the
action written in terms of the Hamiltonian on phase space $(q, p)$ we get Hamilton's equations and
the Hamilton picture
(which in the mock quantum case becomes the Born-Heisenberg-Jordan (matrix) picture)
\be
H \define p \dot{q} - L, \quad 
\delta \Big(\int L\Big) 
=\delta \Big(\int p\,\rd q - H(q, p)\rd t\Big) =0 
\to \dot{q} = \frac{\partial H}{\partial p}, \quad \dot{p}= -\frac{\partial H}{\partial q}.
\ee
From the definition of the action we immediately get the Hamilton-Jacobi picture (which in the quantum case becomes the Schr\"{o}dinger picture/representation)
\be
\frac{ \partial S}{ \partial q} = p, \quad \frac{ \partial S}{ \partial t} 
= - H \Big(q,\, p \define \frac{ \partial S}{ \partial q}\Big).
\ee
Note that the above partial derivatives are with respect to the final position and final time. These equations were the starting point for our discussion in Section~\ref{s:2}.
(However, we could have started from a different point of view,
given the fact that in quantum theory the Lagrangian picture becomes the path integral picture, the Hamiltonian picture becomes the Born-Jordan-Heisenberg
(matrix) picture, and the Hamilton-Jacobi picture becomes the Schr\"{o}dinger (wavefunction) picture.)

Now, our claim is that an emergent quantum dynamics is possible in a complex adaptive environment. Why would that be the case?
In what follows we argue that the robustness of the emergent quantum picture follows already from the above variational principle.

This is easiest to see from the (mock) quantum variational principle (mock Schwinger's variational principle) formally as follows.
The very definition of $\delta$ in the above summary of classical dynamics implies the existence of other ``virtual'' trajectories. What the existence of the canonical quantum constant ($\hbar$) and the non-universal, system-dependent mock Planck constant $\mockbar$, imply is the following
deformed dynamical equation (written with mock quantum theory in mind)
$\delta S = i \mockbar \delta$.
Thus, the $\mockbar \to 0$ limit recovers the above classical equation 
$\delta S =0$, where the action $S$ is still given via the classical expression $S = \int p\,\rd q - H\,\rd t = \int L\,\rd t$.
The non-universal, system-dependent, mock Planck constant $\mockbar$ still carries the units of the action, and the variation derivative $\delta$ is the same on the both sides of the
equation. Finally, one picks as the constant of proportionality $i$ because of the ``democracy'' between all paths, between an initial
and a final point. The classical path is selected by $\delta S=0$, but a definition of $\delta$ requires considering virtual paths as
well. Thus, mock quantum theory brings back all virtual paths/histories of the (adaptive complex) system under consideration and treats them all (including the classical one) on the same footing, provided the complex system is in an adaptive environment that allows for such deformed dynamics.

The above (mock) quantum (Schwinger's) variational principle 
$\delta S = i \mockbar \delta$,
immediately gives the fundamental equations of quantum theory.
Let us pick as a variation derivative, for example, the variation with respect to time $t$. Then by applying the above formal equation to
some functions $\psi(q)$ 
\be
(\delta_t S)\,\psi =( i \mockbar \delta_t)\,\psi 
 ~\to~ 
H\psi = i \mockbar \delta_t \psi,
\ee
where we have used that the time derivative of $S$ with respect to the initial time is $\frac{ \partial S}{ \partial t} =  H $.
(The lower limit in the integral $S = \int p\,\rd q - H\rd t$ is responsible for this sign reversal.)
Thus we obtain the mock Schr\"{o}dinger equation.
Looking instead at the variation with respect to $q$ (initial) yields
\be
(\delta_q S)\,\psi =( i \mockbar \delta_q)\,\psi 
 ~\to~ 
p \psi = - i \mockbar \delta_q \psi,
\ee
because the partial derivative of $S$ with respect to the initial $q$ is $\frac{ \partial S}{ \partial q} = -p$.
In the Hamiltonian operator $H$ one must substitute 
$H(q,\, p\,{=}\,{-}i\mockbar \delta_q)$ and so the Schr\"{o}dinger equation reads
more precisely
\be
H\big(q,\, p\,{=}\,{-}i\mockbar \delta_q\big) \psi 
= i \mockbar \delta_t \psi.
\ee
One may ask: why do we have the formal equation $\delta S = i \mockbar \delta$?
The answer is that this equation is equivalent to Feynman's path integral formulation of mock quantum theory.
According to this formulation the expectation value of any observable $\cO$ is given
by this expression  (written in analogy with equilibrium statistical mechanics)
\be
\langle \cO \rangle = \frac{\int \rD q~ \cO e^{\frac{i S}{\mockbar}}}{\int\rD q~ e^{\frac{i S}{\mockbar}}}.
\ee
Now, this is a complex number, and its variation is zero
\be
\delta \langle \cO \rangle = 0 ~~\to~~
\frac{\int\rD q~ (\delta \cO + \frac{i}{\mockbar} \cO\, \delta S )\,e^{\frac{i S}{\mockbar}}}{\int \rD q~ e^{\frac{i S}{\mockbar}}} =0.
\ee
Therefore we have
\be
\Big\langle\big(\delta \cO + \frac{i}{\mockbar} \cO\, \delta S\big)\Big\rangle =0 
~~\to~~ i \mockbar \langle \delta \cO \rangle = \langle \cO\, \delta S \rangle,
\ee
or by putting $\cO = 1$ we get our original (but more precise) formal version of the (mock) Schwinger variational principle $\langle \delta S \cO \rangle= i \mockbar \langle \delta \cO \rangle$,
which we now understand as a differential form of the path integral.
As we have already shown, this in turn leads to the (mock) Schr\"{o}dinger equation as well as the canonical expression for the momentum operator.

Finally, the most general form of the path integral is the phase space one, from which the coordinate version of the
path integral comes about after integration over the momentum variables
\be
\int\rD q\,\rD p~ e^{i S/\mockbar} \define 
\int\rD q\,\rD p~ e^{\frac{i }{\mockbar} \int (p\rd q - H(q,p)\rd t)} 
 ~\to~ 
\int \rD q~e^{\frac{i }{\mockbar} \int L( q, \dot{q})\rd t}.
\ee
Once one has the (mock) Schr\"{o}dinger equation (derived from the path integral)
one can derive the Born-Heisenberg-Jordan equations of motion --- the Hamiltonian formulation of quantum theory.
One can also derive the  Born-Heisenberg-Jordan equations of motion directly from the Schwinger variational principle, or from the path integral, because 
the path integral is just the Green's function for the Schrodinger operator $i \mockbar \delta_t -H$. (The path integral also follows from the emergent unitary evolution, 
$\psi (t) = \exp\big({-}\frac{i}{\mockbar}\!\int\!H \rd t\big)\,\psi(0)$,
implied by the mock Schr\"{o}dinger equation.)

So, indeed, the analogue of the Hamilton-Jacobi formulation 
is the Schr\"{o}dinger equation, the analogue of the Hamilton formulation is the Born-Heisenberg-Jordan operatorial formulation,
and the path integral is the analogue of the Lagrangian formulation because it is covariant, since 
$S = \int L\,\rd t$ is a Lorentz scalar. 
And all these different ``pictures,'' or representations, are equivalent.
In mock (analog, or emergent) quantum theory, one thereby retains the canonical quantum descriptions but replaces
the canonical, universal, Planck constant with an emergent, or analog, non-universal and system-dependent, mock Planck constant,
$\hbar \to \mockbar$.
The Born rule and the corresponding interpretation of $|\psi|^2$
for the emergent (mock) quantum probability should thus be understood as a steady-state distribution for the non-equilibrium dynamics of the underlying classical system in the presence of a complex adaptive environment that leads to such an emergent quantum description.

\section{Quantum hydrodynamics in a nutshell}
\label{a:C}
To derive 'mock quantum' Euler equation and 'mock quantum' stress-tensor, we start with the de-Broglie-Bohm formulation of classical Hamilton-Jacobi equations
that include $V_Q$ (for the canonical classical Hamiltonian $H = \frac{{\vec{p}}^2}{2m} +V$, with $\nabla S \define \vec{p}$\,)
\be
\frac{\partial S}{\partial t} + \frac{(\nabla S)^2}{2m} +V + V_Q=0.
\ee
Let us take the gradient of this de-Broglie-Bohm equation (act with $\nabla$ from the left side)
\be
\nabla \Big[\frac{\partial S}{\partial t} + \frac{(\nabla S)^2}{2m} +V + V_Q\Big]=0,
\ee
which can be rewritten as
\be
 \Big[\frac{\partial }{\partial t} + \frac{1}{m}(\nabla S \cdot \nabla)\Big] \nabla S = -\nabla[V + V_Q].
\ee
Now, let us use $\vec{v} \define \frac{\nabla S}{m}$ (because in our case $\vec{p} = m \vec{v}$)
to rewrite the previous equation as
\be
\frac{\partial \vec{v}}{\partial t} + (\vec{v} \cdot \nabla) \vec{v} = -\frac{1}{m} \nabla[V + V_Q].
\label{(v.d)v}
\ee
This looks like Euler's equation of classical hydrodynamics (with the extra ``quantum force'' provided by the gradient of the quantum potential $-\nabla V_Q$), 
where the other, continuity equation, is 
given by the conservation of probability (remember, $\psi = \sqrt{\rho}\, e^{iS/{\mockbar}}$)
\be
\frac{\partial \rho}{\partial t} + \nabla (\rho \vec{v}) 
= 0.
\ee
This produces the (mock) quantum Euler equation,
\be
\frac{\partial v_i}{\partial t} + (\vec{v} \cdot \nabla) v_i = -\frac{1}{m} \partial_i V - \frac{1}{m} \partial_i V_Q,
\ee
to be compared with the classical Euler equation that includes the pressure $p$ term (with the fluid density $m \rho$),
\be
\frac{\partial v_i}{\partial t} + (\vec{v} \cdot \nabla) v_i = -\frac{1}{m} \partial_i V - \frac{1}{m \rho} \partial_j (p\, \delta_{ij}).
\ee
Such ``(mock) quantum hydrodynamics'' is defined by the continuity equation $\frac{\partial \rho}{\partial t} + \nabla (\rho \vec{v}) = 0$ and
the ``(mock) quantum Euler'' equation
\be
\frac{\partial v_i}{\partial t} + (\vec{v} \cdot \nabla) v_i = -\frac{1}{m} \partial_i V - \frac{1}{m \rho} \partial_j \sigma_{ij},
\tag{\ref{sig.ij}$'$}
\ee
where the ``(mock) quantum'' stress-tensor $\sigma_{ij}$ reads
\be
\sigma_{ij} \define 
-\frac{\mockbar^2 \rho}{4m}\,\partial_i\partial_j \log{\rho}.
\tag{\ref{s.ij}$'$}
\ee
Once again, as in our previous discussions, in the case of Mock Quantum Theory, we have the same hydrodynamic formulation known from canonical quantum theory, in which 
the Planck constant is replaced by an effective, mock Planck 
constant.


\begin{thebibliography}{25}
\raggedright\frenchspacing

\bibitem{es}
E.~Schr\"{o}dinger,
{\it What is life},
Cambridge 1944.

\bibitem{quantumbiology}
For a review of quantum biology see, for example, A. Marais et al., 
Journal of the Royal Society, 15 (2018) 148.
For a critical view of 
quantum biology consult, for example, J. Cao et al., Science Advances, 16 (2020) 14.

\bibitem{plotnitsky}
A. Plotnitsky, 
Phys. Scr. {\bf T63}, 014011 (19pp) (2014); See also:
A. Plotnitsky and
E. Haven, ed.
{\it The Quantum-Like Revolution}, Springer 2023.

\bibitem{khrenikov}
A. Khrennikov, I. Basieva, E. M. Pothos, and I. Yamato, Scientific Reports, {\bf 8}: 16225 (2018).
See also: 
M Asano, A Khrennikov, M Ohya, Y Tanaka, I Yamato,
{\it Quantum adaptivity in biology: from genetics to cognition},
Springer, 2015,
as well as,
A.~Y.~Khrennikov and 
E.~Haven,
Journal of Mathematical Psychology 53 (5), 378-388 (2009);
E.~Conte, 
A.~Y.~ Khrennikov, O.~Todarello, 
A.~Federici and
L.~Mendolicchio,
Open Systems and Information Dynamics 16 (01), 85-100 (2009).

\bibitem{Minic:2014zsa}
D.~Minic and S.~Pajevic,
Mod. Phys. Lett. B \textbf{30}, no.11, 1650201 (2016)
[arXiv:1409.7588 [quant-ph]].

\bibitem{wen}
For example, X-G. Wen, {\it Quantum Field Theory of Many-body Systems: From the Origin of Sound to an Origin of Light and Electrons },
Oxford, 2007.

\bibitem{tony}
A. J. Leggett, J. Phys.: Condens. Matter {\bf 14}  R415 (2002).


\bibitem{droplet}
Y.~Couder and E.~Fort, 
Phys. Rev. Lett. {\bf 97} 154101 (2006). 

\bibitem{fluidpilot}
J.W.M.~Bush, Annu. Rev. Fluid Mech. {\bf 47} 269–92 (2015)
J.W.M.~Bush, Phys. Today {\bf 68} 47 (2015), and original references therein.

\bibitem{droplet2}
S.~Perrard et al., Nature Comm. {\bf 5} 3219 (2014).

\bibitem{Bush-Oza}
J.W.M.~Bush and A.U.Oza, Rep. Prog. Phys. {\bf 84} 017001 (2021).


\bibitem{eqbook}
L. de la Pena, A. M. Cetto, A. V. Hernandez, {\it The Emerging Quantum: The Physics behind Quantum Mechanics},
Springer, 2015.

\bibitem{anderson}
The classic reference is P.W. Anderson, Science {\bf 177} (4047) 393 (1972).


\bibitem{qcqi}
See for example, Michael A. Nielsen, Isaac L. Chuang, {\it Quantum Computation and Quantum Information},
Cambridge University Press, 2000.

\bibitem{jhholland}
John H Holland,  Journal of Systems Science and Complexity. 19 (1): 1–8
(2006).

\bibitem{bohr}
Our proposal can be related to some old ideas
of Bohr and of Delbruck about the relevance of fundamental frameworks in physics to theoretical biology,
N. Bohr,  Nature, {\bf 131} 421 and 457 (1933); Naturwissensch., {\bf 50} 725 (1963). 
M. Delbr\"{u}ck,
{\it A physicist looks at biology}, Trans. Conn. Acam. Arts and Sci,
reprinted in Cairns, Stent, and Watson, eds, {\it Phage and
the Origins of Molecular Biology}, Cold Spring Harbor, N.Y., (1966);
Science, {\bf 168} 1312 (1970); also in Lex Prix Nobel en 1969, Stockholm,
The Nobel Foundation:
{\it Light and Life III}, Carlsberg, Res. Commun., {\bf 41} 299 (1976). 

\bibitem{thooft}
G. 't Hooft, arXiv:0707.4568v1. 
't Hooft has summarized his work on this topic in  arXiv:1405.1548, devoted to the cellular automaton interpretation of quantum mechanics:
{\it The Cellular Automaton Interpretation of Quantum Mechanics},
Springer, 2016.

\bibitem{davies}
P. Davies, Physics Today, August 2020, 34-40, and references therein.

\bibitem{droplet_critique}
T.~Bohr, A.P.~Andersen and B.~Lautrup, 
in {\it Recent Advances in Fluid Dynamics with Environmental Applications}, Springer, pp. 335-349
\bibitem{droplet_critique2}
G.~Pucci et al., 
J. Fluid Mech. {\bf 835} 1136 (2018).


\bibitem{footnote1}
This cancellation may be thought of as a mutual interaction between the considered system and its environment, without necessarily implying that the behavior of one {\em causes} the behavior of the other.

\bibitem{rs}
R. Schiller, Phys. Rev. {\bf 125} 1100 (1961),
Phys. Rev. {\bf 125} 1109 (1961);
N. Rosen, Am. J. Phys. {\bf 32} 597 (1964), Am. J. Phys. {\bf 33} 146 (1964); 
Found. Phys. {\bf 16} 687 (1985).

\bibitem{lv}
A. J. Lotka, Proc. Natl. Acam. Sci U.S.A {\bf 6} 410 (1920);
J. Amer. Chem. Soc {\bf 42} 1595 (1920);
V. Volterra, Mem. Acad. Lincei {\bf 2} 31 (1926);
{\it Lecons sur la theories mathematique de la lutte pour la vie},
Gauthiers-Villars, Paris, 1931.
There are other equations of the
same nature, such as the Wilson-Cowan model in neuroscience 
H.R. Wilson, J.D. Cowan, 
 Biophys. J. 
{\bf 12} 1 (1972).

\bibitem{other}
Other refs on the variational principle for the Lotka-Volterra equations:
for example, E. H. Kerner, J. Math. Phys, {\bf 38} 1218 (1997). See also:
M. Plank, J. Math. Phys. {\bf 36} 3520 (1995).
Duarte, P., Fernandes, L.R., Oliva, W.M., 
J. Diff. Eqs. {\bf 149} 143 (1998).

\bibitem{holland}
P.R. Holland, {\it The Quantum Theory of Motion}, Cambridge, 1993.

\bibitem{bohm}
D. Bohm and B. J. Hiley, {\it The Undivided Universe: An Ontological Interpretation of Quantum theory}, Routledge, 1995.

\bibitem{rpf}
For an interesting view on the quantum potential consult lecture 21 on superconductivity from The Feynman Lectures on Physics,
volume 3: R. P. Feynman, R. B. Leighton and M. Sands, {\it  The Feynman Lectures on Physics, Quantum Mechanics}, Section~21--8, Addison-Wesley, 1965.


\bibitem{stability}
For the issue of deterministic and stochastic stability in Lotka-Volterra systems see,
A. N. Kolmogorov, Giorn. Instituto Ital. Attuari, {\bf 7} 74 (1936);
M. Mobilia, I. T. Georgiev and U. Tauber, Jour. Stat. Phys.,
{\bf 128} 447 (2007). 
See also: U. Tauber, 
J. Phys. A: Math. Theor. 45 (2012) 405002.

\bibitem{kelso2012}
J. A. S. Kelso, 
Phil.
Tran. of the Roy. Soc. B: Biological Sciences 367.1591 (2012): 906-918.


\bibitem{pisarchik}
A.N. Pisarchik, U. Feudel, 
 Phys. Rep. {\bf 540} 167 (2014).

\bibitem{zurek}
For a review see, W. H. Zurek, Rev. Mod. Phys. {\bf 75} 715 (2003).

\bibitem{bassi}
The role of the environmental noise in the inverse problem of the emergence of classical
theory from the more fundamental quantum dynamics has been reviewed in 
A.  Bassi and G. C. Ghirardi, Phys. Rept. {\bf 379} 257 (2003).


\bibitem{wootters}
W. K. Wootters, Phys. Rev. {\bf D 23} 357 (1981). See also: S. L. Braunstein and
C. M. Caves, Phys. Rev. Lett. {\bf 72} 3439 (1994).

\bibitem{chia}
D. Minic and C.-H. Tze, Phys.Lett. {\bf B 581} 111 (2004).
See also Section~2 and appendix B of V. Jejjala, M. Kavic and D. Minic,
Int. J. Mod. Phys. {\bf A 22} 3317 (2007).
This view is developed in
a novel approach to quantum gravity,  L.~Freidel, R.~G.~Leigh and D.~Minic,
  Phys. Lett. B {\bf 730}, 302 (2014), 
Int. J. Mod. Phys. D \textbf{23}, no.12, 1442006 (2014)
[arXiv:1405.3949 [hep-th]]; 
JHEP \textbf{06}, 006 (2015)
J. Phys. Conf. Ser. \textbf{804}, no.1, 012032 (2017); 
Int. J. Mod. Phys. A \textbf{34}, no.28, 1941004 (2019).

\bibitem{othereq}
For other approaches to the idea of emergent quantum theory, consult for example S.  L. Adler: {\it Quantum Theory as an Emergent Phenomenon: The Statistical Mechanics of Matrix Models as the Precursor of Quantum Field Theory}, Cambridge University Press, 2004. 



\bibitem{thooft1}
In arXiv:1405.1548, 't Hooft points out that by considering a discrete cellular automaton and the discrete time
evolution in terms of {\it permutations} of states of the cellular automaton, one can generate, in an appropriate continuum
temporal limit,
an effective Schr\"{o}dinger equation.


\bibitem{cellular}
J. von Neumann, {\it Theory of Self-Reproducing Automata},  A. Burks, ed., University of Illinois Press, Urbana, IL, 1966;
S. Wolfram, Rev. of Mod. Phys. {\bf 55} 601 (1983);
S. Wolfram, {\it A New Kind of Science}, Wolfram Media, 2002,
\url{https://www.wolframscience.com/nks/}.




\bibitem{hiley}
B. J. Hiley, M. A. de Gosson, G. Dennis,  arXiv:1610.07130 [quant-ph].

\bibitem{sorkin}
R.~D.~Sorkin,
Mod. Phys. Lett. A \textbf{9}, 3119-3128 (1994)
[arXiv:gr-qc/9401003 [gr-qc]].


\bibitem{SRB=BRS}
``Sinai–Ruelle–Bowen measure,''
 \url{https://en.wikipedia.org/wiki/Sinai–Ruelle–Bowen_measure},
 last accessed Feb.~3, 2024;
 Y.G.~Sinai,
 R. Math. Surv. \textbf{27} no.~4 (1972) 21--69;
 R.E.~Bowen,
  pp. 63--76 in {\em Equilibrium states and the ergodic theory of Anosov diffeomorphisms,}
  {\em\/Lecture Notes in Mathematics,} vol.~470 (Springer, 1975);
 D.~Ruelle, 
 Am. J. Math. \textbf{98} no.~3 (1976) 619–654.

\bibitem{Ruelle:2004con}
D.~Ruelle, 
  {\em Physics Today} {\bfseries 57} no.~5, (May, 2004) 48--53.

\bibitem{Curtright:2011vw}
T.L.~Curtright and C.K.~Zachos,
Asia Pac. Phys. Newslett. \textbf{1}, 37-46 (2012),
[arXiv:1104.5269 [physics.hist-ph]].

\bibitem{metabolism}
P.~Hanggi, P.~Talkner and M.~Borkovec,
Rev. Mod. Phys. 62(2) (1990) 251.

\bibitem{WC}
H.R.~Wilson and J.D.~Cowan,
Kybernetik. \textbf{13} no.~2 (1973) 55--80.


\bibitem{latham}
A. Lerchner and P. E. Latham,
``A unifying framework for understanding state-dependent network
dynamics in cortex,'' 
 arXiv:1511.00411 [q-bio.NC].

\bibitem{michel}
M. Henkel and M. Pleimling, {\it Non-equilibrium phase transitions}, volume 2: {\it Ageing and
dynamical scaling far from equilibrium}, Springer, 2010.
See also: U.~C.~T\"{a}uber, {\it Critical Dynamics}, Cambridge, 2014.


\bibitem{smolin}
L.~Smolin,
Found. Phys. \textbf{46}, no.6, 736-758 (2016)
[arXiv:1506.02938 [quant-ph]].
L.~Smolin,
``Views, variety and quantum mechanics,''
[arXiv:2105.03539 [quant-ph]].

\bibitem{barbour}
J.~Barbour and L.~Smolin,
``Extremal variety as the foundation of a cosmological quantum theory,''
[arXiv:hep-th/9203041 [hep-th]].

\bibitem{madelung1927}
E.~Madelung, ``Quantum theory in hydrodynamical form,'' {\em Zeit. f. Phys.}
  {\bfseries 40} (1927) 322. Translated by D.H. Delphenich.


\bibitem{Gray:2013rv}
A.~N.~Kolmogorov, Dokl. Akad. Nauk SSSR. 31: 99–101 (1941). The classic textbook on the topic of turbulence is
U.~Frisch, {\it Turbulence}, Cambridge, 1995.
See also: for example, N.~Gray, D.~Minic and M.~Pleimling,
Int. J. Mod. Phys. A \textbf{28}, 1330009 (2013)
[arXiv:1301.6368 [hep-th]], Section~6, and references therein.
For a recent discussion of the comparison between
theory and experiment:
K.~P.~Iyer et al., Phys. Rev. Lett. 126, 254501 (2021).

\bibitem{cellosc}
M. Maroto, N. Monk, {\it Cellular oscillatory mechanisms}. Vol. 641. Springer Science \& Business Media, 2008.

\bibitem{synthetic}
M. B. Elowitz  and S. Leibler, 
Nature 403.6767 (2000): 335-338;
J. Stricker, S. Cookson, M. R. Bennett, W. H. Mather, L.S. Tsimring and J. Hasty, 
Nature,456 (7221), 516-519 (2008);
A. Prindle, J. Selimkhanov, H. Li, I. Razinkov, L. S. Tsimring, and J. Hasty, Nature , 508, 387-391 (2014);
W. Mather, J. Hasty, and L. S. Tsimring,
 Phys. Rev. Lett., 113, 128102 (2014).

\bibitem{enz}
C. P. Enz, Foundations of Physics, {\bf 24} 1281 (1994).

\bibitem{parisi}
G. Parisi and Wu, Sci. Sin, {\bf 24} 484 (1981); P. H. Damgaard and H. Huffel,
Phys. Rep. {\bf 152} 227 
(1987). See  also: 
E.~Nelson, {\it Dynamical Theories of Brownian Motion}, Princeton, 1966; {\it Quantum Fluctuations}, Princeton, 1985.

\bibitem{quantumbio}
R, Dorner, J. Goold, L. Heaney, T. Farrow, P. G. Roberts, J. Hirst, V. Vedral,
 arXiv:1111.1646 and references therein.

\bibitem{supermq}
Apart from emergent mock quantum theory, stochastic complex dynamics might lead,
to emergent mock super-quantum or gravitized quantum theories. Super-quantum theories are discussed in
 L.~N.~Chang, Z.~Lewis, D.~Minic and T.~Takeuchi,
  J.\ Phys.\ A {\bf 46}, 485306 (2013),  L.~N.~Chang, Z.~Lewis, D.~Minic and T.~Takeuchi,
  arXiv:1312.0645 [quant-ph],  L.~N.~Chang, Z.~Lewis, D.~Minic, T.~Takeuchi and C.~-H.~Tze,
  Adv.\ High Energy Phys.\  {\bf 2011}, 593423 (2011). See also: 
  S. Popescu, D. Rohrlich, Found. Phys. 24 (3) (1994) 379–385;
S. Popescu, Nat. Phys. 10 (4) (2014) 264–270, arXiv:1012.5810.
For gravitized quantum theory consult, 
L.~Freidel, R.~G.~Leigh and D.~Minic,
Phys. Rev. D \textbf{94}, no.10, 104052 (2016)
[arXiv:1606.01829 [hep-th]];
D.~Minic,
[arXiv:2003.00318 [hep-th]], in the proceedings of the
10th Mathematical Physics Meeting : School and Conference on Modern Mathematical Physics,
9-14 September 2019, Belgrade, Serbia, (C19-09-09.8);
P.~Berglund, T.~H\"ubsch, D.~Mattingly and D.~Minic,
Int. J. Mod. Phys. D \textbf{31}, no.14, 2242024 (2022);
P.~Berglund, A.~Geraci, T.~H\"ubsch, D.~Mattingly and D.~Minic,
Class. Quant. Grav. \textbf{40}, no.15, 155008 (2023).
See also: R. Penrose, Foundations of Physics, 44, 557 (2014);
L. Hardy, [arXiv:quant-ph/0101012 [quant-ph]]; [arXiv:gr-qc/0509120 [gr-qc]].

\bibitem{bialek}
W.~Bialek, {\it Biophysics: Searching for Principles}, Princeton, 2012.

\bibitem{feynman}
R.~P.~Feynman and A.~R.~Hibbs, {\it Quantum Mechanics and Path Integrals}, McGraw-Hill, 1965.

\end{thebibliography}
\end{document}